\documentclass[prc,preprint,
  showpacs,showkeys,lengthcheck,
  nofootinbib,tightenlines,onecolumn,notitlepage,
  preprintnumbers,superscriptaddress]{revtex4-1}
\usepackage[utf8]{inputenc}
\usepackage{newtxtext}
\usepackage[mathcal]{euscript}

\usepackage[export]{adjustbox}
\usepackage{float}
\usepackage{lipsum}
\usepackage{fixltx2e}

\usepackage{graphicx}
\usepackage[caption=false]{subfig}
\usepackage[usenames,dvipsnames]{xcolor}
\graphicspath{{figures/}}

\usepackage{tikz}
\usetikzlibrary{arrows.meta,decorations.pathmorphing,backgrounds,positioning,fit,petri,shapes.geometric}

\usepackage{hyperref}
\hypersetup{colorlinks=true,citecolor=blue,linkcolor=blue,urlcolor=blue}

\usepackage{diagbox}
\usepackage{booktabs}

\usepackage{amsmath,amssymb,amsfonts}
\usepackage{mathrsfs}
\usepackage{slashed}
\usepackage{bm}

\newcommand{\eq}[1]{Eq.~(\ref{#1})}

\def\barx{{\bar x}}\def\l{\lambda}
\def\bfz{{\bf z}}\def\r{\rho}\def\cM{{\cal M}}
\def\D{\Delta}\newcommand{\bfkappa}{\mbox{\boldmath $\kappa$}}\def\k{\kappa}\def\bfP{{\bf P}}
\def\bfk{{\bf k}}
\def\k{\kappa}
\def\bfb{{\bf b}}\def\bfr{{\bf r}}\def\bfs{{\bf s}}\def\bfR{{\bf R}}
\newcommand{\bea}{\begin{eqnarray}}
\newcommand{\eea}{\end{eqnarray}}
\newcommand{\bfDelta}{\mbox{\boldmath $\Delta$}}\def\a{\alpha}

\begin{document}

\title{
  Unified formalism for electromagnetic and gravitational probes:
  densities
}

\author{Adam Freese}
\email{afreese@uw.edu}
\address{Department of Physics, University of Washington, Seattle, WA 98195, USA}

\author{Gerald A. Miller}
\email{miller@uw.edu}
\address{Department of Physics, University of Washington, Seattle, WA 98195, USA}

\begin{abstract}
  The use of light front coordinates allows a fully relativistic description
  of a hadron's spatial densities to be obtained.
  These densities must be two-dimensional and transverse to a chosen spatial direction.
  We explore their relationship to the three-dimensional,
  non-relativistic densities,
  with a focus on densities associated with the energy momentum tensor.
  The two-dimensional non-relativistic densities can be obtained from the light front densities
  through a non-relativistic limit,
  and can subsequently be transformed into three-dimensional non-relativistic
  densities through an inverse Abel transform.
  However, this operation is not invertible,
  and moreover the application of the inverse Abel transform to the light front densities
  does not produce a physically meaningful result.
  We additionally find that the Abel transforms of so-called Breit-frame
  densities  
  generally differ significantly from the true
  light front densities.
  Numerical examples are provided to illustrate the various differences
  between the light front, Breit frame, and non-relativistic treatment of densities.
\end{abstract}

\preprint{NT@UW-21-09}

\maketitle


\section{Introduction}

The energy momentum tensor (EMT)
and the associated gravitational form factors~\cite{Kobzarev:1962wt}
have recently attracted significant interest in the hadron physics community.
Major open questions such as the
proton mass puzzle~\cite{Ji:1994av,Ji:1995sv,Lorce:2017xzd,Hatta:2018sqd}
and proton spin puzzle~\cite{Ashman:1987hv,Ji:1996ek,Leader:2013jra}
are directly related to the EMT.
Moreover, the EMT encodes information about the magnitude and distribution
of forces within
hadrons~\cite{Polyakov:2002yz,Polyakov:2018zvc,Lorce:2018egm,Freese:2021czn},
a topic which has itself led to a flurry of
theoretical studies~\cite{Polyakov:2018zvc,Lorce:2018egm},
empirical extractions~\cite{Burkert:2018bqq,Dutrieux:2021nlz,Burkert:2021ith},
and lattice calculations~\cite{Shanahan:2018nnv}.

\begin{figure}
  \usetikzlibrary{positioning}
  \begin{tikzpicture}

    \tikzstyle{elsa} = [
      rectangle, rounded corners,
      minimum width=3cm, minimum height=1cm, inner sep=0.25cm,
      text centered,
      draw=black, fill=RoyalBlue!30
    ]
    \tikzstyle{lightfront} = [
      diamond,
      minimum width=3cm, minimum height=1cm,
      text centered,
      draw=black, fill=green!30
    ]
    \tikzstyle{breit} = [
      ellipse,
      minimum width=3cm, minimum height=1cm,
      text centered,
      draw=black, fill=red!30
    ]
    \tikzstyle{nonrel} = [
      rectangle,
      minimum width=3cm, minimum height=1cm,
      text centered,
      draw=black, fill=Orange!30
    ]

    \tikzstyle{arrow} = [thick,->,>=stealth,draw=Violet]
    \tikzstyle{arrow2} = [thick,<->,>=stealth,draw=ForestGreen]
    \tikzstyle{arrow0} = [thick,dotted,>=stealth,draw=BrickRed]

    \node[elsa] (gtmd) {$H(x,\mathbf{k}_\perp,\xi,\boldsymbol{\Delta}_\perp;Q^2)$};
    \node[elsa] (cff) [below=of gtmd] {$\mathcal{H}(\xi,\boldsymbol{\Delta}_\perp^2;Q^2)$};
    \node[elsa] (gpd) [left=3.2cm of cff] {$H(x,\xi,\boldsymbol{\Delta}_\perp^2;Q^2)$};
    \node[elsa] (exp) [right=3.2cm of cff] {EIC, JLab};

    \node[elsa] (gff) [below=of gpd] {
      $F(\boldsymbol{\Delta}_\perp^2)$,
      $A(\boldsymbol{\Delta}_\perp^2)$,
      $D(\boldsymbol{\Delta}_\perp^2)$
    };
  
    \node[lightfront] (lf) [right=3.2cm of gff] {$\rho_{\mathrm{LF}}(\mathbf{b}_\perp)$};
  
    \node[breit] (bf3d) [below=1.7cm of gff] {$\rho_{\mathrm{BF}}^{(\mathrm{3D})}(\mathbf{r})$};
    \node[breit] (bf2d) [right=3.2cm of bf3d] {$\rho_{\mathrm{BF}}^{(\mathrm{2D})}(\mathbf{b}_\perp)$};
  
    \node[nonrel] (nr3d) [below=of bf2d] {$\rho_{\mathrm{NR}}^{(\mathrm{3D})}(\mathbf{r})$};
    \node[nonrel] (nr2d) [right=3.2cm of nr3d] {$\rho_{\mathrm{NR}}^{(\mathrm{2D})}(\mathbf{b}_\perp)$};
  
    \draw [arrow] (gtmd) -- node[color=Violet!70!black, anchor=south east ] {$\int \mathrm{d}^2\mathbf{k}_\perp$} (gpd);
    \draw [arrow] (gpd)  -- node[color=Violet!70!black, anchor=west] {$\int \mathrm{d}x \, x^{n-1}$} (gff);
    \draw [arrow] (bf2d) -- node[color=Violet!70!black, anchor=south west] {FNR} (nr2d);
    \draw [arrow] (bf3d) -- node[color=Violet!70!black, anchor=south west] {FNR} (nr3d);
    \draw [arrow] (lf)   -- node[color=Violet!70!black, anchor=south west] {FNR} (nr2d);
    \draw [arrow] (gff)  -- node[color=Violet!70!black, 
      anchor=west, text width=1.7cm, align=center
    ] {Breit frame Fourier transform} (bf3d);
    \draw [arrow] (gpd)   -- node[
      color=Violet!70!black, anchor=south, text width=2.8cm, align=center
    ] {$\int \mathrm{d}x \, C(x,\xi)$} (cff);

    \draw [arrow2] (gff)  -- node[anchor=south, text width=3cm, align=center, color=ForestGreen!70!black] {Fourier transform} (lf);
    \draw [arrow2] (bf3d) -- node[anchor=south, text width=3cm, align=center, color=ForestGreen!70!black] {Abel transform} (bf2d);
    \draw [arrow2] (nr3d) -- node[anchor=south, text width=3cm, align=center, color=ForestGreen!70!black] {Abel transform} (nr2d);
    \draw [arrow2] (cff) -- node[anchor=south, text width=1.7cm, align=center, color=ForestGreen!70!black] {Experiment} (exp);

    \draw [arrow0] (lf) -- node[anchor=south east,text=BrickRed] {???} (bf2d);

    \node (text1) [above=0cm of gtmd] {GTMDs};
    \node (text2) [left=0.25cm of gpd] {GPDs};
    \node (text3) [left=0.25cm of gff] {FFs};
    \node (text4) [above=0cm of cff] {Compton form factors};

  \end{tikzpicture}
  \caption{
    Chart depicting the relationships between densities in
    different formalisms,
    as well as their relationships to GPDs.
    Here, $\rho$ stands in for any space-dependent density,
    including mass density, $P^+$ density, or even pressure density.
    One-arrow (purple) lines signify one-way relationships,
    and two-arrow (green) lines signify invertible relationships.
    There is no systematic connection between light front
    and Breit frame densities.
  }
  \label{fig:chart}
\end{figure}
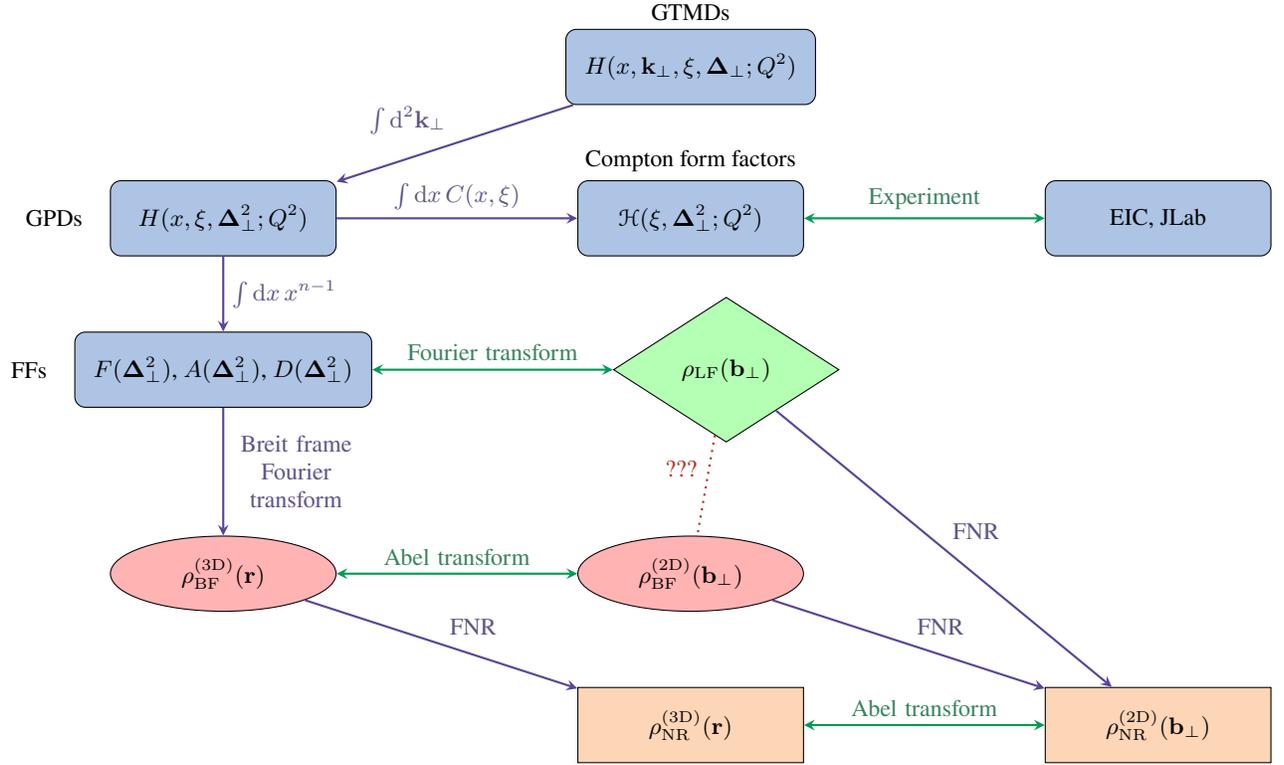

The theoretical studies are driven by the promise of making relevant experiments
to determine the various matrix elements that allow the extraction of
the relevant form factors.
As depicted in Fig.~\ref{fig:chart},
the relevant formalism is most generally expressed through
generalized transverse momentum distributions,
which are obtained from bilocal correlation function
$H_\Gamma(k,P,\D)$ by integrating over $k^-$.
For quarks, this correlator is given by~\cite{Diehl:2015uka}:
\begin{align}
  H_\Gamma(k,P,\D)
  =
  \frac{1}{(2\pi)^{4}}
  \int \mathrm{d}^4z\,
  e^{ik\cdot z}
  \left\langle P+{1\over2}\D\middle|
  \bar q\left(-{z\over2}\right)
  \Gamma
  \mathcal{W}\left(-\frac{z}{2},\frac{z}{2}\right)
  q\left({z\over2}\right)
  \middle|P-{1\over2}\D\right\rangle
  \,,
\end{align}
where $\Gamma$ stands in for a matrix in the Dirac algebra
(e.g., $\gamma^+$)
and $\mathcal{W}(y,x)$ is a Wilson line from $x$ to $y$.
Integration over $\mathbf{k}_\perp$ gives the generalized parton distribution,
Mellin moments of which encode local form factors of interest---including
those appearing in the EMT.

The form factors appearing in matrix elements of the EMT encode spatial
densities via Fourier transforms.
When performing these Fourier transforms,
it is important to keep perspective about the actual, physical meaning
of the densities that are obtained.
It has been established~\cite{Burkardt:2002hr,Miller:2007uy,Miller:2009sg,Miller:2018ybm,Freese:2021czn}
that the only meaningful way to obtain fully relativistic densities
is through two-dimensional Fourier transforms at fixed light front time.
The three-dimensional Breit frame density
is obtained by erroneously assuming that the hadron can be spatially
localized~\cite{Miller:2018ybm,Jaffe:2020ebz}.
Nonetheless, densities obtained through three-dimensional Fourier transforms
unfortunately remain ubiquitous in the EMT density literature.

The Abel transform has recently been proposed as a means of connecting
the light front and Breit frame formalisms~\cite{Panteleeva:2021iip,Kim:2021jjf}.
It is therefore necessary to explore the meaning of this connection.
The Abel transform can be obtained by integrating one coordinate of a
spherically symmetric density.
However, there is no manifest spherical symmetry on the light front~\cite{Brodsky:1997de}.
Additionally, Refs.~\cite{Panteleeva:2021iip,Kim:2021jjf}
looked at the case of spin-half hadrons, but not spin-zero hadrons,
where the proposed connection is shown below not to work.

The purpose here is to explore the actual meaning of 3D EMT densities
and their relationship to the fully relativistic 2D light front densities.
In particular, we show that physically meaningful 3D densities can be defined only
in a non-relativistic approximation,
either by taking $c\rightarrow\infty$ or---in some cases---keeping up
to order $v^2/c^2$ corrections.
Additionally, we examine the physical meaning and applicability
of the Abel transform.

This work is organized as follows.
In Sec.~\ref{sec:abel},
we discuss the Abel transform and when it does and does not
connect 2D and 3D densities.
In Sec.~\ref{sec:densities},
we consider the formalism for relativistic and non-relativistic densities
for both spin-zero and spin-half particles,
deriving results for the relationships between them.
Numerical examples, based on using a simple hadronic
model~\cite{Miller:2009sg,Weinberg:1966jm,Gunion:1973ex} are used
to study the implications of using the Breit frame and the non-relativistic limit
in Sec.~\ref{sec:model}.
We conclude and provide a summary in Sec.~\ref{sec:conclusion}.

\section{Abel transforms of physical densities}
\label{sec:abel}

Since the fully relativistic light front densities are two-dimensional,
they can be most directly compared to two-dimensional
rather than three-dimensional non-relativistic densities.
The 2D non-relativistic densities are obtained by
integrating out one coordinate of a given three-dimensional density
$\rho^{(3D)}_{\rm NR} (\bfr)$
(which may obtained as a three-dimensional Fourier transform of a form factor):
\begin{align}
  \label{eqn:abel:gen}
  \rho_{\mathrm{NR}}^{(2D)}(\mathbf{b}_\perp)
  =
  \int_{-\infty}^\infty \mathrm{d}z \,
  \rho_{\mathrm{NR}}^{(3D)}(\mathbf{r})
  \,,
\end{align}
where $\mathbf{b}_\perp = (x,y)$ are the transverse coordinates.
If the 3D density is spherically symmetric,
i.e.,
 $\rho_{\mathrm{NR}}^{(3D)}(\mathbf{r})$ is a function of only $r=|\mathbf{r}|$,
 a change of integration variable allows us to write:
\begin{align}
  \label{eqn:abel}
  \rho_{\mathrm{NR}}^{(2D)}(b_\perp)
  =
  2 \int_b^\infty \mathrm{d}r \,
  \frac{r}{\sqrt{r^2-b_\perp^2}}
  \rho_{\mathrm{NR}}^{(3D)}(r)
  \equiv
  \mathscr{A}\Big[
    \rho_{\mathrm{NR}}^{(3D)}(r)
    \Big](b_\perp)
  \,,
\end{align}
which defines the Abel transform\footnote{
  The Abel transform has several slightly
  different definitions in the literature.
  Eq.~(\ref{eqn:abel}) agrees with the definition in
  Ref.~\cite{Bracewell:2000abl},
  which we use here because of its clear geometrical meaning.
  Ref.~\cite{Panteleeva:2021iip} uses a different definition of the Abel transform.
}.
For densities which depend on individual components of $\mathbf{r}$,
one must use Eq.~(\ref{eqn:abel:gen}).
However, each of the densities we consider can be written in terms of
derivatives of an entirely scalar density
to which Eq.~(\ref{eqn:abel}) can be applied.

One pertinent property of the Abel transform is that
it is invertible~\cite{Bracewell:2000abl}:
\begin{align}
  \label{eqn:abel:inv}
  \rho_{\mathrm{NR}}^{(3D)}(r)
  =
  - \frac{1}{\pi}
  \int_r^\infty \mathrm{d}b_\perp \,
  \frac{
    \rho_{\mathrm{NR}}^{(2D)}(b_\perp)
  }{\mathrm{d}b_\perp}
  \frac{1}{\sqrt{b_\perp^2-r^2}}
  \equiv
  \mathscr{A}^{-1}\Big[
    \rho_{\mathrm{NR}}^{(2D)}(b_\perp)
    \Big](r)
  \,.
\end{align}
This allows a 3D non-relativistic density to be reconstructed from
a 2D non-relativistic density,
assuming that we know the former to be spherically symmetric ahead of time.

The importance of spherical symmetry cannot be stressed enough.
If one begins with an azimuthally symmetric 2D density without a guarantee
of spherical symmetry in three dimensions,
applying Eq.~(\ref{eqn:abel:inv}) may not give the correct 3D density.
Consider, for instance, the following 3D densities:
\begin{subequations}
  \begin{align}
    f(r)
    &=
    \frac{a^3}{(x^2+y^2+z^2+a^2)^3}
    \,,
    \\
    g(\mathbf{r})
    &=
    \frac{sa^3}{(x^2+y^2+s^2z^2+a^2)^3}
    \,,
  \end{align}
\end{subequations}
where $a$ is some length scale and $s$ is a positive unitless constant.
These both integrate to the same azimuthally symmetric function:
\begin{align}
  \int_{-\infty}^\infty \mathrm{d}z\, 
  f(r)
  =
  \int_{-\infty}^\infty \mathrm{d}z\, 
  g(\mathbf{r})
  =
  \mathscr{A}\big[f(r)\big](b)
  \equiv
  F(b)\,=\,{3a^3\pi\over 8 (a^2+b^2)^{5/2}}
  \,.
\end{align}
Applying the inverse Abel transform to $F(b)$ will return $f(r)$,
even if---in the context of a physical scenario---$g(\mathbf{r})$
is the true 3D density.

This point is especially pertinent since the light front
Galilean subgroup of the Poincar\'e group does not have an
$\mathrm{SO}(3)$ subgroup.
Light front dynamics does not admit 3D spherical symmetry~\cite{Brodsky:1997de},
so it is meaningless to try to construct an exact relativistic 3D density by
applying Eq.~(\ref{eqn:abel:inv}) to a light front density,
as done in Refs.~\cite{Panteleeva:2021iip,Kim:2021jjf}
or earlier in Ref.~\cite{Rajan:2018zzy}.
In fact, there are model calculations suggesting that the proton
is elongated in the $x^-$ direction~\cite{Miller:2019ysh}.
Moreover, we shall see below that densities of
transversely-polarized hadrons have $\phi$ dependence,
demonstrating that
spherical symmetry in $(\mathbf{x}_\perp,x^-)$ does not hold.
At best, the inverse Abel transform of a light front density
can give the 3D density in a non-relativistic approximation,
as we shall show below.

One helpful property of Eq.~(\ref{eqn:abel:gen})
that will aid the exploration to follow
is its effect on Fourier transforms.
If a 3D density is defined by:
\begin{align}
  \rho_{\mathrm{NR}}^{(3D)}(r)
  =
  \int \frac{\mathrm{d}^3\boldsymbol{\Delta}}{(2\pi)^3}
  F(t=-\boldsymbol{\Delta}^2)
  e^{-i\boldsymbol{\Delta}\cdot\mathbf{r}}
  \,,
\end{align}
then because the $z$ integral of $e^{-i\Delta_z z}$ is
$2\pi\delta(\Delta_z)$, one has:
\begin{align}
  \label{eqn:abel:fourier}
  \rho_{\mathrm{NR}}^{(2D)}(b_\perp)
  =
  \int \frac{\mathrm{d}^2\boldsymbol{\Delta}_\perp}{(2\pi)^2}
  F(t=-\boldsymbol{\Delta}_\perp^2)
  e^{-i\boldsymbol{\Delta}_\perp\cdot\mathbf{b}_\perp}
  \,.
\end{align}


\section{Relativistic and non-relativistic densities of the EMT}
\label{sec:densities}

We shall now consider relativistic and non-relativistic densities
of the energy momentum tensor (EMT).
These densities are related to form factors,
which are defined via matrix elements of plane wave states.
For spin-zero particles, the standard decomposition is:
is~\cite{Polyakov:2018zvc}:
\begin{align}
  \label{eqn:emt:spin0}
  \langle p' |
  T^{\mu\nu}(0)
  |p\rangle
  =
  2 P^\mu P^\nu A(t)
  + \frac{\Delta^\mu\Delta^\nu-\Delta^2g^{\mu\nu}}{2} D(t)
  \,,
\end{align}
while for spin-half particles~\cite{Polyakov:2018zvc}:
\begin{align}
  \langle p',\lambda |
  T^{\mu\nu}(0)
  |p,\lambda\rangle
  =
  \bar{u}(p',\lambda) \left\{
    \frac{P^\mu P^\nu}{M} A(t)
    +
    \frac{\Delta^\mu\Delta^\nu-\Delta^2g^{\mu\nu}}{4M} D(t)
    +
    \frac{iP^{\{\mu}\sigma^{\nu\}\rho}\Delta_\rho}{2M} J(t)
    \right\}
  u(p,\lambda)
  \,,
\end{align}
where $P = \frac{1}{2}\big(p+p'\big)$,
$\Delta = p'-p$,
$t = \Delta^2$,
and curly brackets $\{\}$ signify symmetrization, e.g.,
$a^{\{\mu}b^{\nu\}} = a^\mu b^\nu + a^\nu b^\mu$.


\subsection{Relativistic light front densities}
\label{sec:densities:lf}

As discussed in Refs.~\cite{Brodsky:1997de,
Burkardt:2002hr,Miller:2007uy},
the only way to meaningfully define relativistic densities is at fixed
light front time,
since this allows separation between barycentric and relative coordinates.
For both spin-zero hadrons and \emph{longitudinally polarized}
spin-half hadrons,
the light front momentum ($P^+$) density is found to
be~\cite{Freese:2021czn}:
\begin{align}
  \label{eqn:p+}
  \rho_{P^+}^{(\mathrm{LF})}(\mathbf{b}_\perp)
  =
  P^+
  \int \frac{\mathrm{d}^2\boldsymbol{\Delta}_\perp}{(2\pi)^2}
  A(t)
  e^{-i\boldsymbol{\Delta}_\perp\cdot\mathbf{b}_\perp}
  \,,
\end{align}
and in these same cases, the comoving stress tensor
is~\cite{Lorce:2018egm,Freese:2021czn}:
\begin{align}
  \label{eqn:Sij:LF}
  S_{\mathrm{LF}}^{ij}(\mathbf{b}_\perp)
  =
  \frac{1}{4P^+}
  \Big(
  \delta^{ij} \boldsymbol{\nabla}_\perp^2
  -
  \boldsymbol{\nabla}_\perp^i \boldsymbol{\nabla}_\perp^j
  \Big)
  \int \frac{\mathrm{d}^2\boldsymbol{\Delta}_\perp}{(2\pi)^2}
  D(t)
  e^{-i\boldsymbol{\Delta}_\perp\cdot\mathbf{b}_\perp}
  \,.
\end{align}
This tensor can be decomposed into a isotropic pressure $p^{\mathrm{LF}(b_\perp)}$
and pressure anisotropy (or shear stress function) $s^{\mathrm{LF}(b_\perp)}(b_\perp)$:
\begin{align}
  \label{eqn:Sij:LF:ps}
  S_{\mathrm{LF}}^{ij}(\mathbf{b}_\perp)
  =
  \delta^{ij} p^{(\mathrm{LF})}(b_\perp)
  +
  \left( \frac{b_\perp^ib_\perp^j}{b_\perp^2} - \frac{1}{2}\delta^{ij} \right)
  s^{(\mathrm{LF})}(b_\perp)
  \,,
\end{align}
and has eigenpressures in the tangential
directions~\cite{Lorce:2018egm,Panteleeva:2021iip,Freese:2021qtb}:
\begin{subequations}
  \label{eqn:prt:LF}
  \begin{align}
    \label{eqn:pr:LF}
    p^{(\mathrm{LF})}_r(b_\perp)
    &=
    \hat{b}_i \hat{b}_j S_{\mathrm{LF}}^{ij}(\mathbf{b}_\perp)
    =
    p^{(\mathrm{LF})}(b_\perp) + \frac{1}{2} s^{(\mathrm{LF})}(b_\perp)
    \\
    \label{eqn:pr:LF}
    p^{(\mathrm{LF})}_t(b_\perp)
    &=
    \hat{\phi}_i \hat{\phi}_j S_{\mathrm{LF}}^{ij}(\mathbf{b}_\perp)
    =
    p^{(\mathrm{LF})}(b_\perp) - \frac{1}{2} s^{(\mathrm{LF})}(b_\perp)
    \,.
  \end{align}
\end{subequations}

A useful quantity is the potential $\widetilde{D}(b_\perp)$, defined by:
\begin{align}
  \label{eqn:Dtilde}
  \widetilde{D}(b_\perp)
  &=
  \frac{ 1 }{4P^+}
  \int \frac{\mathrm{d}^2\boldsymbol{\Delta}_\perp}{(2\pi)^2}
  D(t)
  e^{-i\boldsymbol{\Delta}_\perp\cdot\mathbf{b}_\perp}
  \,.
\end{align}
The comoving stress tensor is related to this potential by:
\begin{align}
  S^{ij}(\mathbf{b}_\perp)
  &=
  \Big( \nabla_i \nabla_j - \delta_{ij} \nabla^2 \Big)
  \widetilde{D}(b_\perp)
  \,.
\end{align}
The radial and tangential pressures have simple expressions in terms of
the potential:
\begin{subequations}
  \begin{align}
    p_r(b_\perp)
    &=
    \frac{1}{b_\perp}
    \frac{\mathrm{d}\widetilde{D}(b_\perp)}{\mathrm{d}b_\perp}\label{pr}
    \\
    p_t(b_\perp)
    &=
    \frac{\mathrm{d}^2\widetilde{D}(b_\perp)}{\mathrm{d}b_\perp^2}
    \,.
  \end{align}
\end{subequations}


\subsubsection{Transversely polarized hadrons}

It's possible to prepare spin-half hadrons in transversely polarized states,
for which the light front densities will no longer have azimuthal symmetry.
A transversely-polarized hadron can be prepared as a superposition of
light front helicity states~\cite{Carlson:2008zc}:
\begin{align}
  | s_T = \mathbf{s}_\perp \rangle
  =
  \frac{
    \left| \lambda = +1 \right\rangle
    +
    e^{i\phi_s}
    \left| \lambda = -1 \right\rangle
  }{ \sqrt{2} }
  \,.
\end{align}
In terms of helicity states, matrix elements of transversely polarized states
take the form (with momentum dependence suppressed to compactify the formula):
\begin{align}
  \langle \mathbf{s}_\perp | \hat{O} | \mathbf{s}_\perp \rangle
  =
  \frac{1}{2} \Big\{
    \langle + | \hat{O} | + \rangle
    +
    \langle - | \hat{O} | - \rangle
    +
    \langle + | \hat{O} | - \rangle
    e^{i\phi_s}
    +
    \langle - | \hat{O} | + \rangle
    e^{-i\phi_s}
    \Big\}
  \,,
\end{align}
which is the average of helicity state densities,
plus an additional $\phi$-dependent helicity-flip contribution.

The $P^+$ density of transversely polarized states is:
\begin{align}
  \label{eqn:p+:trans}
  \rho_{P^+,T}^{(\mathrm{LF})}(\mathbf{b}_\perp,\mathbf{s}_\perp)
  =
  \rho_{P^+}^{(\mathrm{LF})}(b_\perp)
  +
  P^+ \,
  \frac{\sin(\phi)}{2Mc}
  \frac{\mathrm{d}}{\mathrm{d}b_\perp}
  \int \frac{\mathrm{d}^2\boldsymbol{\Delta}_\perp}{(2\pi)^2}
  \Big( A(t) - 2J(t) \Big)
  e^{-i\boldsymbol{\Delta}_\perp\cdot\mathbf{b}_\perp}
  \,,
\end{align}
where $\rho_{P^+}^{(\mathrm{LF})}(b_\perp)$ is density for helicity states
(equal to the spin-zero $P^+$ density in Eq.~(\ref{eqn:p+})),
and $\phi = \phi_b - \phi_s$ is the angle
from the transverse polarization vector to the transverse coordinate.

It's worth remarking that the angular dependence is a strictly relativistic effect because 
taking the $c\rightarrow\infty$ limit eliminates the $\phi$ dependence.  

The comoving stress tensor of also has angular dependence for
transversely polarized states, and in addition has a new tensorial structure:
\begin{subequations}
  \begin{align}
    S^{ij}_T(\mathbf{b}_\perp,\mathbf{s}_\perp)
    &=
    \delta_{ij}
    p_T(b_\perp,\phi)
    +
    \left( \hat{b}^i\hat{b}^j - \frac{1}{2}\delta^{ij} \right)
    s_T(b_\perp,\phi)
    +
    \big( \hat{b}^i \hat{\phi}^j + \hat{b}^j \hat{\phi}^i \big)
    v_T(b_\perp)
    \\
    p_T(b_\perp,\phi)
    &=
    p(b_\perp)
    +
    \frac{\sin(\phi)}{2Mc}
    p'(b_\perp)
    \\
    s_T(b_\perp,\phi)
    &=
    s(b_\perp)
    +
    \frac{\sin(\phi)}{2Mc}
    s'(b_\perp)
    \\
    v_T(b_\perp,\phi)
    &=
    \frac{\cos(\phi)}{2Mc}
    \frac{s(b_\perp)}{b_\perp}
    \,,
  \end{align}
\end{subequations}
where $p(b_\perp)$ and $s(b_\perp)$ are the isotropic pressure and anisotropy
functions in the helicity state case,
and where $\hat{b}$ and $\hat{\phi}$ are unit vectors.
The quantities $p_T,\,s_T,\,v_T$ cannot be obtained through an Abel transform.

The new tensor structure associated with $v_T(b_\perp,\phi)$ is peculiar
and does not contribute to either the radial or the tangential pressure,
since it contracts with both $\hat{b}^i\hat{b}^j$ and with
$\hat{\phi}^i\hat{\phi}^j$ to zero.
It is more instructive---as discussed in Ref.~\cite{Polyakov:2018zvc}---to
find the eigenvalues and eigenvectors of the comoving stress tensor.
Since the eigenvectors satisfy:
\begin{align}
  S^{ij}_T(\mathbf{b}_\perp,\mathbf{s}_\perp) \hat{e}_{\pm}^j
  =
  p_\pm(b_\perp,\phi)
  \hat{e}_{\pm}^i
  \,,
\end{align}
where we use $\pm$ to index the two eigenvectors
(and their associated eigenpressures),
it is possible to write the pressure in any direction as a superposition
of the eigenpressures.
If we write the eigenvectors in terms of an angle with respect to the
transverse spin vector:
\begin{align}
  \hat{e}_\pm
  =
  \cos(\theta_\pm)\mathbf{s}_\perp + \sin(\theta_\pm)\tilde{\mathbf{s}}_\perp
  \,,
\end{align}
then the eigenvectors of $S^{ij}_T(\mathbf{b}_\perp,\mathbf{s}_\perp)$ are
given by the angles:
\begin{subequations}
  \begin{align}
    \theta_+
    &=
    \phi
    + \frac{1}{2}\tan^{-1}\left(\frac{2v_T(b_\perp,\phi)}{s_T(b_\perp,\phi)}\right)
    \\
    \theta_-
    &=
    \phi
    + \frac{1}{2}\tan^{-1}\left(\frac{2v_T(b_\perp,\phi)}{s_T(b_\perp,\phi)}\right)
    + \frac{\pi}{2}
  \end{align}
  while the associated eigenpressures are given by:
  \begin{align}
    p_\pm(b_\perp,\phi)
    &=
    p_T(b_\perp,\phi)
    \pm
    \sqrt{
      \frac{1}{4} \big(s_T(b_\perp,\phi)\big)^2 + \big(v_T(b_\perp,\phi)\big)^2
    }
    \,.
  \end{align}
\end{subequations}
In the limit of large $b_\perp$,
one has $v_T(b_\perp,\phi) \ll s_T(b_\perp,\phi)$,
so the eigen-angles become $\phi$ and $\phi + \frac{\pi}{2}$
very far from the center of the hadron.
This deformation from the radial and tangential directions
be seen as a relativistic effect
that vanishes in the $c\rightarrow\infty$ limit,
along with the angular dependence of the eigenpressures.

The angular dependence in both the $P^+$ density and the stress tensor
demonstrates a lack of spherical symmetry in the light front formalism.
This is of course not surprising, since rotations around the $x$ and $y$ axes
are dynamical operators that do not commute with
the light front Hamiltonian $P^-$~\cite{Dirac:1949cp,Brodsky:1997de}.
This finding precludes use of the inverse Abel transform to construct a
physically meaningful 3D relativistic density.
Moreover, the inverse Abel transform cannot even be applied at a formal level,
since the transform acts on an azimuthally symmetric function
of a single variable.


\subsection{Breit frame densities}
\label{sec:densities:bf}

If one tries to define 3D relativistic densities at fixed instant form time,
the density becomes contaminated by center-of-mass motion of the hadron
as a whole~\cite{Jaffe:2020ebz}.
It is controversial whether localization of the center-of-mass motion
in coordinate space is relativistically possible
(see Refs.~\cite{Newton:1949cq,Kalnay:1967zz,Pavsic:2018vbs} for attempts, however),
and localization in momentum space produces infinite radii for all
densities~\cite{Miller:2018ybm}
owing to the Heisenberg uncertainty principle.

Nonetheless, ostensibly relativistic 3D densities are
ubiquitous throughout the hadron physics literature.
The so-called Breit frame densities are \emph{defined}
by taking a Fourier transform of e.g. Eq.~(\ref{eqn:emt:spin0})
with respect to the momentum transfer
$\boldsymbol{\Delta}$ while setting the average momentum $\mathbf{P}=0$.
It is worth stressing that these densities have not been derived
from the basic definition of a physical density,
i.e., expected value of a local current for a physical hadron state.
As shown in Ref.~\cite{Miller:2018ybm},
the original derivation in Ref.~\cite{Sachs:1962zzc} was erroneous
and neglected a term that would make all radii infinite.
Nonetheless, the erroneous Breit frame densities with finite radii
are ubiquitous enough that they should be addressed,
despite not being physically meaningful relativistic densities.

The Breit frame mass density and stress tensor both have different
expressions for spin-zero and spin-half particles.
For the mass density~\cite{Polyakov:2002yz,Hudson:2017xug,Polyakov:2018zvc}:
\begin{subequations}
  \begin{align}
    \rho_{\mathrm{mass}}^{(j=0)}(\mathbf{r})
    &=
    M
    \int \frac{\mathrm{d}^3\boldsymbol{\Delta}}{(2\pi)^3}
    \frac{1}{\sqrt{1 - \frac{t}{4M^2}}}
    \left\{
      A(t)
      -
      \frac{t}{4M}
      \Big[ A(t) + D(t) \Big]
      \right\}
    e^{-i\boldsymbol{\Delta}\cdot\mathbf{r}}
    \\
    \label{eqn:mass:BF:half}
    \rho_{\mathrm{mass}}^{(j=1/2)}(\mathbf{r})
    &=
    M
    \int \frac{\mathrm{d}^3\boldsymbol{\Delta}}{(2\pi)^3}
    \left\{
      A(t)
      -
      \frac{t}{4M}
      \Big[ A(t) + D(t) - 2J(t) \Big]
      \right\}
    e^{-i\boldsymbol{\Delta}\cdot\mathbf{r}}
    \,,
  \end{align}
\end{subequations}
while for the stress tensor:
\begin{subequations}
  \begin{align}
    T^{ij}_{\mathrm{BF},0}(\mathbf{r})
    &=
    \frac{1}{4M}
    \Big(
    \delta^{ij} \boldsymbol{\nabla}_\perp^2
    -
    \boldsymbol{\nabla}_\perp^i \boldsymbol{\nabla}_\perp^j
    \Big)
    \int \frac{\mathrm{d}^3\boldsymbol{\Delta}}{(2\pi)^3}
    \frac{1}{\sqrt{1-\frac{t}{4M^2}}}
    D(t)
    e^{-i\boldsymbol{\Delta}\cdot\mathbf{r}}
    \\
    T^{ij}_{\mathrm{BF},\frac{1}{2}}(\mathbf{r})
    &=
    \frac{1}{4M}
    \Big(
    \delta^{ij} \boldsymbol{\nabla}_\perp^2
    -
    \boldsymbol{\nabla}_\perp^i \boldsymbol{\nabla}_\perp^j
    \Big)
    \int \frac{\mathrm{d}^3\boldsymbol{\Delta}}{(2\pi)^3}
    D(t)
    e^{-i\boldsymbol{\Delta}\cdot\mathbf{r}}
  \end{align}
\end{subequations}
Using Eq.~(\ref{eqn:abel:fourier}), it is possible to obtain simple
formulas for the two-dimensional reductions of these Breit frame
stress tensors.
We find:
\begin{subequations}
  \begin{align}
    T^{ij}_{\mathrm{BF},0}(\mathbf{b}_\perp)
    &=
    \frac{1}{4M}
    \Big(
    \delta^{ij} \boldsymbol{\nabla}_\perp^2
    -
    \boldsymbol{\nabla}_\perp^i \boldsymbol{\nabla}_\perp^j
    \Big)
    \int \frac{\mathrm{d}^2\boldsymbol{\Delta}_\perp}{(2\pi)^2}
    \frac{1}{\sqrt{1-\frac{t}{4M^2}}}
    D(t)
    e^{-i\boldsymbol{\Delta}_\perp\cdot\mathbf{b}_\perp}
    \\
    \label{eqn:Sij:BF:2D}
    T^{ij}_{\mathrm{BF},\frac{1}{2}}(\mathbf{b}_\perp)
    &=
    \frac{1}{4M}
    \Big(
    \delta^{ij} \boldsymbol{\nabla}_\perp^2
    -
    \boldsymbol{\nabla}_\perp^i \boldsymbol{\nabla}_\perp^j
    \Big)
    \int \frac{\mathrm{d}^2\boldsymbol{\Delta}_\perp}{(2\pi)^2}
    D(t)
    e^{-i\boldsymbol{\Delta}_\perp\cdot\mathbf{b}_\perp}
    \,,
  \end{align}
\end{subequations}
where the use of $\mathbf{b}_\perp$ instead of $\mathbf{r}$
is used to signify that these are 2D functions.

Comparing Eqs.~(\ref{eqn:Sij:LF},\ref{eqn:Sij:BF:2D}),
one can see that
$P^+ S^{ij}_{\mathrm{LF}}(\mathbf{b}_\perp)
=
M T^{ij}_{\mathrm{BF},\frac{1}{2}}(\mathbf{b}_\perp)$, i.e.,
that the 2D Breit frame and light front comoving stress tensors
have identical forms (up to a constant) for spin-half particles specifically.
This is the essentially the central finding of Ref.~\cite{Panteleeva:2021iip}.
The Abel transform connects 2D Breit frame densities to 3D Breit frame
densities, and it just so happens
that the Breit frame and light front comoving stress tensors have similar
integrands specifically for spin-half particles.
Because of this, Abel transforms \emph{formally} work out to relate
the 3D Breit frame and 2D light front pressures for spin-half particles.
It should be recalled however that the light front does not have 3D
spherical symmetry and that the Breit frame densities are not physically
meaningful densities. Thus 
 this formal coincidence does not have any deep physical meaning,
and does not lend  credence to the Breit frame pressure.

By contrast, one can easily observe that
$P^+ S^{ij}_{\mathrm{LF}}(\mathbf{b}_\perp)
\neq
M T^{ij}_{\mathrm{BF},0}(\mathbf{b}_\perp)$,
so the findings of Ref.~\cite{Panteleeva:2021iip} do not apply to
spin-zero particles.
3D Breit frame pressures in spin-zero hadrons are not related to
2D light front pressures by Abel transforms.
In light of the caveats we have stressed so far,
this is not surprising,
but it does help stress that the findings of
Ref.~\cite{Panteleeva:2021iip} originate from a
coincidence rather than a deep connection between the light front and Breit frame.


\subsection{Non-relativistic densities}
\label{sec:nr}

In a non-relativistic (NR) quantum mechanical theory,
just as in relativistic quantum field theory,
the density associated with a local operator $\hat{\mathcal{O}}(x)$
and a physical state $|\Psi\rangle$ is given by:
\begin{align}
  \rho_{\mathrm{NR}}(\mathbf{r})
  =
  \langle\Psi|\hat{\mathcal{O}}(\mathbf{r})|\Psi\rangle_{\mathrm{NR}}
  \,.
\end{align}
The meaning of the term NR is that the system obeys Galilean invariance,
in which the dependence on relative and center-of mass variables can be separated.
The center of mass position of the physical state $|\Psi\rangle$
generally has a finite spatial extent.
This state can be localized by allowing the total momentum to have infinite extent. 
This localization can be achieved, for example,
by using a Gaussian representation~\cite{Miller:2018ybm} so that:
\begin{align}
  \label{eqn:delta}
  \Psi(\mathbf{p},\mathbf{s})
  =
  (2\pi)^{3/4} (2\sigma)^{3/2} e^{-\sigma^2 \mathbf{p}^2}
  \,,
\end{align}
in which $\mathbf{p}$ refers to the total momentum of the system, 
and then taking the $\sigma\rightarrow0$ limit at the end of the calculation.
The spatial dependence of $\rho_{\mathrm{NR}}(\mathbf{r})$ thus defined will
encode only internal structure of the hadron.
It is possible to show
(using similar derivations to those in Refs.~\cite{Miller:2018ybm,Freese:2021czn})
that the density can be written:
\begin{align}
  \label{eqn:fourier:nr}
  \rho_{\mathrm{NR}}(\mathbf{r})
  =
  \lim_{\sigma\rightarrow0}
  \,
  (2\pi)^{3/2} (2\sigma)^3
  \int \frac{\mathrm{d}^3\mathbf{P}}{(2\pi)^3}
  \int \frac{\mathrm{d}^3\boldsymbol{\Delta}}{(2\pi)^3}
  e^{-2\sigma^2\mathbf{P}^2}
  \langle \mathbf{p}',\mathbf{s} | \hat{\mathcal{O}}(0) | \mathbf{p},\mathbf{s} \rangle_{\mathrm{NR}}
  e^{-\frac{\sigma^2}{2}\boldsymbol{\Delta}^2}
  e^{-i\boldsymbol{\Delta}\cdot\mathbf{r}}
  \,
\end{align}
in which $\mathbf{P}=\frac{1}{2}\big(\mathbf{p}+\mathbf{p}'\big)$
and $\boldsymbol{\Delta} = \mathbf{p}' - \mathbf{p}$.
The limit $\sigma\rightarrow0$ is to be taken after
the $\mathbf{P}$ integral has been done.
Integrals in which the matrix element contains factors
$\mathbf{P}^2$, $\mathbf{P}^4$, etc.\ will diverge,
which limits the densities that can be considered;
for instance, we cannot calculate a kinetic energy density
for a completely spatially localized system.


\subsubsection{Non-relativistic reduction}

The matrix element
$\langle \mathbf{p}',\mathbf{s} | \hat{\mathcal{O}}(0) | \mathbf{p},\mathbf{s} \rangle_{\mathrm{NR}}$
appearing in Eq.~(\ref{eqn:fourier:nr}) is a \emph{non-relativistic}
matrix element\footnote{
  Since matrix elements are invariant under unitary transformations,
  unlike state kets or operators,
  it is more suitable to apply non-relativistic reduction to matrix elements
  as a whole rather than to their individual components.
}.
In practice, one knows how to express the relativistic counterpart
$\langle \mathbf{p}',\mathbf{s} | \hat{\mathcal{O}}(0) | \mathbf{p},\mathbf{s} \rangle_{\mathrm{rel}}$
in terms of local form factors.
It should be possible to obtain the former from the latter by restoring factors
of $c$ where appropriate and taking the $c\rightarrow\infty$ limit.
Before doing so, we also must bear in mind that the momentum kets are normalized
differently in the relativistic and non-relativistic cases;
the conventional (instant form) normalization for momentum kets is:
\begin{subequations}
  \begin{align}
    \langle \mathbf{p}',\mathbf{s}' | \mathbf{p},\mathbf{s} \rangle_{\mathrm{NR}}
    &=
    (2\pi)^3 \delta^{(3)}(\mathbf{p}'-\mathbf{p}) \delta_{ss'}
    \\
    \langle \mathbf{p}',\mathbf{s}' | \mathbf{p},\mathbf{s} \rangle_{\mathrm{rel}}
    &=
    (2\pi)^3 (2E_{\mathbf{p}}) \delta^{(3)}(\mathbf{p}'-\mathbf{p}) \delta_{ss'}
    \,.
  \end{align}
\end{subequations}
Thus, the \emph{fully non-relativistic} (FNR) limit is given by:
\begin{align}
  \label{eqn:FNR:element}
  \langle \mathbf{p}',\mathbf{s} | \hat{\mathcal{O}}(0) | \mathbf{p},\mathbf{s} \rangle_{\mathrm{NR}}
  =
  \lim_{c\rightarrow\infty}
  \frac{1}{\sqrt{2E_{\mathbf{p}}2E_{\mathbf{p}'}}}
  \langle \mathbf{p}',\mathbf{s} | \hat{\mathcal{O}}(0) | \mathbf{p},\mathbf{s} \rangle_{\mathrm{rel}}
  \,.
\end{align}
Of course, one can take $E_{\mathbf{p}}/c^2\rightarrow M$ in the FNR limit,
but in some cases it may be instructive to know what the leading relativistic
corrections look like.
These can be found by expanding the RHS of Eq.~(\ref{eqn:FNR:element})
as a power series in $|\mathbf{p}|/(Mc)$, and dropping terms beyond a certain order
instead of taking the $c\rightarrow\infty$ limit.

When taking the non-relativistic limit,
consistency demands that this limit be applied to the whole of
the RHS of Eq.~(\ref{eqn:FNR:element}).
For instance, in the case of the electric charge density
of a spin-zero hadron, one has:
\begin{align}
  \label{eqn:FNR:example}
  \langle \mathbf{p}',\mathbf{s} | j^0(0) | \mathbf{p},\mathbf{s} \rangle_{\mathrm{NR}}
  =
  \lim_{c\rightarrow\infty}
  \left\{
    \frac{
      \big(E_\mathbf{p}+E_{\mathbf{p}'} \big)
    }{\sqrt{2E_{\mathbf{p}}2E_{\mathbf{p}'}}}
    F(t)
    \right\}
  =
  \lim_{c\rightarrow\infty}
  F(t)
  \equiv
  F_{\mathrm{NR}}(t)
  \,.
\end{align}
Since the dynamics that govern the structure of hadrons
are manifestly relativistic,
the form factor $F(t)$ will change in the non-relativistic limit,
as seen for instance in Ref.~\cite{Miller:2009sg}.
Consistent application of the non-relativistic limit thus means
that the form factors appearing in non-relativistic 3D densities
and fully relativistic light front densities should be different
functions.
We shall subscript the latter using NR.


\subsubsection{Non-relativistic mass density}

In the non-relativistic formalism,
matrix elements of $T^{00}$ provide the mass density.
For spin-zero and spin-half particles, respectively, we have:
\begin{subequations}
  \begin{align}
    \label{eqn:T00:spin0}
    \langle \mathbf{p}',\mathbf{s} |
    T^{00}(0)
    |\mathbf{p},\mathbf{s}\rangle_{\mathrm{NR}}^{(j=0)}
    \approx
    Mc^2
    \Bigg\{
      \bigg[
        1
        & + \frac{\mathbf{P}^2}{2M^2c^2}
        + \frac{\boldsymbol{\Delta}^2}{8M^2c^2}
        \bigg]
      A_{\mathrm{NR}}(t)
      +
      \frac{\boldsymbol{\Delta}^2}{4M^2c^2} D_{\mathrm{NR}}(t)
      +
      \mathcal{O}(1/c^4)
      \Bigg\}
    \\
    \langle \mathbf{p}',\mathbf{s} |
    T^{00}(0)
    |\mathbf{p},\mathbf{s}\rangle_{\mathrm{NR}}^{(j=1/2)}
    \approx
    Mc^2
    \Bigg\{
      \bigg[
        1
        & + \frac{\mathbf{P^2}}{2M^2c^2}
        + \frac{\boldsymbol{\Delta}^2}{4M^2c^2}
        - \frac{ i (\boldsymbol{\Delta}\times\mathbf{P})\cdot\mathbf{s} }{ 4M^2c^2 }
        \bigg]
      A_{\mathrm{NR}}(t)
      \notag \\ &
      +
      \frac{\boldsymbol{\Delta}^2}{4M^2c^2} D_{\mathrm{NR}}(t)
      +
      \left[
        - \frac{\boldsymbol{\Delta}^2}{2M^2c^2}
        + \frac{ i (\boldsymbol{\Delta}\times\mathbf{P})\cdot\mathbf{s} }{ M^2c^2 }
        \right]
      J_{\mathrm{NR}}(t)
      +
      \mathcal{O}(1/c^4)
      \Bigg\}
    \,,
  \end{align}
\end{subequations}
where the NR subscripts on the form factors indicate that they should
be expanded in powers of $1/c$ as well,
and where here we use $|\mathbf{s}|=1$ for simplicity.
Because of the $\mathbf{P}^2$ terms, these cannot be used to define
densities for an arbitrarily localized hadron at order $1/c^2$,
though the $c\rightarrow\infty$ limit does not have this issue.
However, if this matrix element is placed into Eq.~(\ref{eqn:fourier:nr})
\emph{without} taking the $\sigma\rightarrow0$ limit,
one obtains results in the form:
\begin{subequations}
  \begin{align}
    \label{eqn:energy:nr}
    \rho_{\mathrm{energy}}(\mathbf{r};\sigma)
    =
    \frac{ \rho_{\mathrm{mass}}(\mathbf{r};\sigma) }{M}
    \left\{
      Mc^2
      +
      \frac{\langle\mathbf{P^2}\rangle_\sigma}{2M}
      \right\}
    +
    \mathcal{O}(1/c^2)
    \,,
  \end{align}
  where:
  \begin{align}
    \langle\mathbf{P}^2\rangle_\sigma
    &=
    \int \frac{\mathrm{d}^3\mathbf{P}}{(2\pi)^3}
    \, \mathbf{P}^2 \,
    |\psi(\mathbf{P},\mathbf{s};\sigma)|^2
    =
    \frac{3}{\sigma^2}
    \\
    \rho_{\mathrm{mass}}^{(j=0)}(\mathbf{r};\sigma)
    &=
    M
    \int \frac{\mathrm{d}^3\boldsymbol{\Delta}}{(2\pi)^3}
    \left\{
      A_{\mathrm{NR}}(t)
      +
      \frac{\boldsymbol{\Delta}^2}{8M^2c^2}
      \Big[
        A_{\mathrm{NR}}(t) + 2D_{\mathrm{NR}}(t)
        \Big]
      \right\}
    e^{-i\boldsymbol{\Delta}\cdot\mathbf{r}}
    e^{-\frac{\sigma^2}{2}\boldsymbol{\Delta}^2}
    +
    \mathcal{O}(1/c^4)
    \\
    \label{eqn:mass:nr:half}
    \rho_{\mathrm{mass}}^{(j=1/2)}(\mathbf{r};\sigma)
    &=
    M
    \int \frac{\mathrm{d}^3\boldsymbol{\Delta}}{(2\pi)^3}
    \left\{
      A_{\mathrm{NR}}(t)
      +
      \frac{\boldsymbol{\Delta}^2}{4Mc^2}
      \Big[ A_{\mathrm{NR}}(t) + D_{\mathrm{NR}}(t) - 2J_{\mathrm{NR}}(t) \Big]
      \right\}
    e^{-i\boldsymbol{\Delta}\cdot\mathbf{r}}
    e^{-\frac{\sigma^2}{2}\boldsymbol{\Delta}^2}
    +
    \mathcal{O}(1/c^4)
  \end{align}
\end{subequations}
This has exactly the expected form of a mass density
plus a (non-relativistic) kinetic energy density.
Although we cannot take the $\sigma\rightarrow0$ limit for the full
(mass+kinetic) energy density at this order in $1/c^2$,
we can actually take this limit for the mass density by itself.
This suggests that we can obtain a meaningful leading-order
relativistic correction to the mass density.
This suggestion must however be tempered by the realization
that the separation of energy into mass and kinetic energy
requires the ability to bring the system to rest,
which is explicitly precluded by taking the $\sigma\rightarrow0$ limit.

It is worthwhile to observe that the NR+LO mass density
for spin-half particles, as given in Eq.~(\ref{eqn:mass:nr:half})
is identical in form to the Breit frame mass density given in
Eq.~(\ref{eqn:mass:BF:half})
(and previously found in Ref.~\cite{Polyakov:2018zvc} for instance).
A caveat worth mentioning is that consistency of the non-relativistic reduction
requires expanding the form factors themselves in powers of $1/c$,
while the Breit frame density uses the exact relativistic form factors.
Moreover, such a coincidence does not occur for spin-zero particles.

The procedure outlined here cannot be used at arbitrarily
high orders in $1/c$, and it is therefore not possible
to define a fully relativistic 3D mass density
through series of relativistic corrections.
The dependence of
$E_{\mathbf{p}}$ and $E_{\mathbf{p}'}$
on
$(\mathbf{P}\cdot\boldsymbol{\Delta})$
prevents factorizing the density integrand into a
$\mathbf{P}$-dependent factor and $\boldsymbol{\Delta}$-dependent factor.

For both spin-zero and spin-half particles,
the fully non-relativistic ($c\rightarrow\infty$) limit gives the same
form for the mass density:
\begin{align}
  \label{eqn:mass:FNR}
  \rho_{\mathrm{mass}}^{(\mathrm{FNR})}(\mathbf{r})
  &=
  M
  \int \frac{\mathrm{d}^3\boldsymbol{\Delta}}{(2\pi)^3}
  A_{\mathrm{NR}}(t)
  e^{-i\boldsymbol{\Delta}\cdot\mathbf{r}}
  \,.
\end{align}
This is spherically symmetric,
and comparison to Eqs.~(\ref{eqn:abel:fourier},\ref{eqn:p+})
shows that this is related to the non-relativistic limit of
the fully relativistic $P^+$ density:
\begin{align}
  \label{eqn:abel:p+}
  \rho_{\mathrm{mass}}^{(\mathrm{FNR})}(r)
  =
  \lim_{c\rightarrow\infty}
  \frac{M}{P^+}
  \mathscr{A}\Big[
    \rho_{P^+}^{(\mathrm{LF})}(b_\perp)
    \Big](r)
  \,.
\end{align}
We thus see that the inverse Abel transform of the $P^+$ density
does have a physical meaning,
if it is accompanied by the $c\rightarrow\infty$ limit:
it gives the \emph{fully non-relativistic} 3D mass density.
Since the $c\rightarrow\infty$ limit is not invertible,
this relation is not invertible either.


\subsubsection{Non-relativistic stress tensor}

The $T^{ij}$ components of the EMT give the stress tensor.
It is worth stressing that $T^{ij}$ does not only encode pressure and shear
forces, but also contains contributions from the motion of the system.
For instance, taking the FNR limit for a spin zero system gives:
\begin{align}
  \langle \mathbf{p}',\mathbf{s} |
  T^{ij}(0)
  |\mathbf{p},\mathbf{s}\rangle_{\mathrm{NR}}^{(j=0)}
  &=
  \frac{\mathbf{P}^i\mathbf{P}^j}{M}
  A_{\mathrm{NR}}(t)
  +
  \left(
  \frac{
    \boldsymbol{\Delta}^i \boldsymbol{\Delta}^j
    -
    \delta_{ij}\boldsymbol{\Delta}^2
  }{4M}
  \right)
  D_{\mathrm{NR}}(t)
  +
  \mathcal{O}(1/c^2)
  \,.
\end{align}
As it is, the stress tensor
cannot be used with Eq.~(\ref{eqn:fourier:nr}) unless the
$\sigma\rightarrow0$ limit is avoided,
because the factor $\mathbf{P}^i\mathbf{P}^j$ multiplying $A(t)$
will produce a $\sigma^{-2}$ divergence when $i=j$.
However, we can define a density at non-zero $\sigma$:
\begin{subequations}
  \label{eqn:stress:FNR:full}
  \begin{align}
    T^{ij}_{\mathrm{NR}}(\mathbf{r};\sigma)
    =
    \rho_{\mathrm{mass}}^{(\mathrm{NR})}(\mathbf{r};\sigma)
    \langle \mathbf{v}^i \mathbf{v}^j \rangle_\sigma
    +
    S^{ij}_{\mathrm{NR}}(\mathbf{r};\sigma)
    \,,
  \end{align}
  in the FNR, where:
  \begin{align}
    \langle \mathbf{v}^i \mathbf{v}^j \rangle_\sigma
    &=
    \int \frac{\mathrm{d}^3\mathbf{P}}{(2\pi)^3}
    \, \frac{\mathbf{P}^i \mathbf{P}^j}{M^2} \,
    |\psi(\mathbf{P},\mathbf{s};\sigma)|^2
    =
    \frac{\delta_{ij}}{M^2\sigma^2}
    \,,
    \\
    S^{ij}_{\mathrm{NR}}(\mathbf{r};\sigma)
    &=
    \int \frac{\mathrm{d}^3\boldsymbol{\Delta}}{(2\pi)^3}
    \left(
    \frac{
      \boldsymbol{\Delta}^i \boldsymbol{\Delta}^j
      -
      \delta_{ij}\boldsymbol{\Delta}^2
    }{4M}
    \right)
    D_{\mathrm{NR}}(t)
    e^{-i\boldsymbol{\Delta}\cdot\mathbf{r}}
    e^{-\frac{\sigma^2}{2}\boldsymbol{\Delta}^2}
    \,.
  \end{align}
\end{subequations}
This has the form expected of the classical
non-relativistic stress tensor,
with a piece encoding movement of the system
and a piece expressing the comoving stress tensor $S^{ij}$.
The comoving stress tensor is invariant under Galilean boosts,
and thus can be interpreted as the stress tensor as seen
by a comoving observer---a physical interpretation that
is justified by having taken the fully non-relativistic limit.

The comoving stress tensor is well-defined in
the $\sigma\rightarrow0$ limit:
\begin{align}
  \label{eqn:Sij:FNR}
  S^{ij}_{\mathrm{NR}}(\mathbf{r})
  =
  \frac{1}{4M}
  \Big(
  \delta^{ij} \boldsymbol{\nabla}^2
  -
  \boldsymbol{\nabla}^i \boldsymbol{\nabla}^j
  \Big)
  \int \frac{\mathrm{d}^3\boldsymbol{\Delta}}{(2\pi)^3}
  D_{\mathrm{NR}}(t)
  e^{-i\boldsymbol{\Delta}\cdot\mathbf{r}}
  \,.
\end{align}
It is straightforward to show that
$S^{ij}_{\mathrm{NR}}(\mathbf{r})$
also has this form for spin-half systems.

The leading relativistic corrections introduce factors of
$\mathbf{P}^2$ and $(\mathbf{P}\cdot\boldsymbol{\Delta})^2$
that preclude using the $\mathcal{O}(1/c^2)$ corrections to define a density
via Eq.~(\ref{eqn:fourier:nr}).
Moreover, outside of the fully non-relativistic limit,
the isolation of a comoving stress tensor is not clear,
and the latter is certainly no longer invariant under boosts.
We thus constrain ourselves to considering
the fully non-relativistic limit for the stress tensor.

As explained in Refs.~\cite{Polyakov:2018zvc,Lorce:2018egm},
this comoving stress tensor can be decomposed into an isotropic pressure
$p(r)$ and a pressure anisotropy $s(r)$:
\begin{align}
  \label{eqn:Sij:ps:FNR}
  S_{\mathrm{NR}}^{ij}(\mathbf{r})
  =
  \delta^{ij} p^{(\mathrm{NR})}(r)
  +
  \left( \frac{r^ir^j}{r^2} - \frac{1}{3}\delta^{ij} \right)
  s^{(\mathrm{NR})}(r)
  \,.
\end{align}
By contracting this with unit vectors, it is possible to obtain
directional pressures, e.g. radial and tangential
pressures~\cite{Polyakov:2018zvc,Lorce:2018egm}:
\begin{subequations}
  \label{eqn:prt:FNR}
  \begin{align}
    \label{eqn:pr:FNR}
    p_r^{(\mathrm{NR})}(r)
    &=
    \hat{r}_i \hat{r}_j S_{\mathrm{NR}}^{ij}(\mathbf{r})
    =
    p^{(\mathrm{NR})}(r) + \frac{2}{3} s^{(\mathrm{NR})}(r)
    \\
    \label{eqn:pt:FNR}
    p^{(\mathrm{NR})}_t(r)
    &=
    \hat{\phi}_i \hat{\phi}_j S_{\mathrm{NR}}^{ij}(\mathbf{r})
    =
    p^{(\mathrm{NR})}(r) - \frac{1}{3} s^{(\mathrm{NR})}(r)
    \,.
  \end{align}
\end{subequations}
By integrating out the $z$ coordinate, one can obtain the 2D
non-relativistic stress tensor:
\begin{align}
  \label{eqn:Sij:FNR:2D}
  S^{ij}_{\mathrm{NR}}(\mathbf{b}_\perp)
  =
  \frac{1}{4M}
  \Big(
  \delta^{ij} \boldsymbol{\nabla}_\perp^2
  -
  \boldsymbol{\nabla}_\perp^i \boldsymbol{\nabla}_\perp^j
  \Big)
  \int \frac{\mathrm{d}^2\boldsymbol{\Delta}_\perp}{(2\pi)^2}
  D_{\mathrm{NR}}(t)
  e^{-i\boldsymbol{\Delta}_\perp\cdot\mathbf{b}_\perp}
  \,,
\end{align}
where dependence on $\mathbf{b}_\perp$ instead of
$\mathbf{r}$ signifies that this is a 2D function.
This can be compared to the light front comoving stress tensor
in Eq.~(\ref{eqn:Sij:LF}), giving:
\begin{align}
  S^{ij}_{\mathrm{NR}}(\mathbf{b}_\perp)
  =
  \lim_{c\rightarrow\infty}
  \frac{P^+}{M}
  S^{ij}_{\mathrm{LF}}(\mathbf{b}_\perp)
  \,.
\end{align}
By comparing the 3D non-relativistic eigenpressures
in Eq.~(\ref{eqn:prt:FNR})
to the 2D light front eigenpressures
in Eq.~(\ref{eqn:prt:LF}),
it is possible also to show that:
\begin{subequations}
  \label{eqn:abel:pressure}
  \begin{align}
    p_t^{(\mathrm{NR})}(r)
    &=
    \lim_{c\rightarrow\infty}
    \frac{P^+}{M}
    \mathscr{A}^{-1}\big[
      p_t^{(\mathrm{LF})}(b_\perp)
      \big](r)
    \\
    p_r^{(\mathrm{NR})}(r)
    &=
    \lim_{c\rightarrow\infty}
    \frac{2P^+}{M}
    \mathscr{A}^{-1}\big[
      p_r^{(\mathrm{LF})}(b_\perp)
      \big](r)
    \,.
  \end{align}
\end{subequations}
These are compatible with the spin-half results of
Ref.~\cite{Panteleeva:2021iip},
although our result applies to spin-zero particles as well.
Just as with the $P^+$ density,
we find that the inverse Abel transform has a physical meaning
when accompanied by the $c\rightarrow\infty$ limit:
it produces the 3D densities in the fully non-relativistic limit.


\subsubsection{Non-relativistic form factors}

The connection between the 3D non-relativistic densities
and the 2D light front densities,
as given in Eqs.~(\ref{eqn:abel:p+},\ref{eqn:abel:pressure})
are not invertible.
This is so because the $c\rightarrow\infty$ limit cannot be undone.
However, if the $c\rightarrow\infty$ limit had no effect on
the form factors---i.e., if
$A_{\mathrm{NR}}(t) = A(t)$
and
$D_{\mathrm{NR}}(t) = D(t)$---then
these relationships would actually be invertible.
However, this is not the case,
and we illustrate these points in the next section using a simple model.


\section{Model calculations}
\label{sec:model}

We will now illustrate the findings of this work with
pedagogical model calculations, specifically using a 
generalization \cite{Miller:2009sg} of the
$\phi^3$ model first used by Weinberg~\cite{Weinberg:1966jm}
and later by by Gunion \textsl{et al.}~\cite{Gunion:1973ex}.
We use the interaction Lagrangian:
\begin{align}
  \mathscr{L}_I[\Psi,\phi,\xi]
  =
  g \Psi(x) \phi(x) \xi(x)
  \,,
\end{align}
where all of the three different fields are spin-zero bosons.
The $\Psi$ particle of mass $M$
represents the bound state of the two different constituents
$\phi$ and $\xi$, of masses $m_1$ and $m_2$ respectively.

The point-like coupling of this model is very simple,
which is a pedagogic advantage,
but short distance effects are emphasized~\cite{Miller:2009sg}
as the light-front wave function has a logarithmic divergence
for small values of the transverse separation $b_\perp$ between the quarks.
Furthermore, the asymptotic behavior of the
electromagnetic form factor is $F(t)\sim {\log^2(-t)\over (-t)}$.

The electromagnetic current of the three-scalar model is given by:
\begin{align}
  \langle p' | j^\mu(0) | p \rangle
  =
  2 P^\mu F(t)
  &=
  i
  g^2e_1
  \int \frac{\mathrm{d}^4k}{(2\pi)^4}
  \frac{
    2 k^\mu
  }{
    \big[ (k-P)^2-m_2^2 \big]
    \big[ (k+\Delta/2)^2-m_1^2 \big]
    \big[ (k-\Delta/2)^2-m_1^2 \big]
  }
  \notag \\ &+
  i
  g^2e_2
  \int \frac{\mathrm{d}^4k}{(2\pi)^4}
  \frac{
    2 k^\mu
  }{
    \big[ (k-P)^2-m_1^2 \big]
    \big[ (k+\Delta/2)^2-m_2^2 \big]
    \big[ (k-\Delta/2)^2-m_2^2 \big]
  }
  \,,
\end{align}
where $P = \frac{1}{2}\big(p+p'\big)$ and $\Delta=p'-p$.
This is a sum of contributions from the constituents with masses $m_1$ and $m_2$.
The $++$ component of the gravitational current is given by:
\begin{align}
  \langle p' | T^{++}(0) | p \rangle
  =
  2 (P^+)^2 A(t)
  &=
  i
  g^2\int \frac{\mathrm{d}^4k}{(2\pi)^4}
  \frac{
    2 (k^+)^2
  }{
    \big[ (k-P)^2-m_2^2 \big]
    \big[ (k+\Delta/2)^2-m_1^2 \big]
    \big[ (k-\Delta/2)^2-m_1^2 \big]
  }
  \notag \\ &+
  ig^2
  \int \frac{\mathrm{d}^4k}{(2\pi)^4}
  \frac{
    2 (k^+)^2
  }{
    \big[ (k-P)^2-m_1^2 \big]
    \big[ (k+\Delta/2)^2-m_2^2 \big]
    \big[ (k-\Delta/2)^2-m_2^2 \big]
  }
  \,,
\end{align}
and to isolate $D(t)$ we look at the $12$ component:
\begin{align}
  \langle p' | T^{12}(0) | p \rangle
  =
   {1\over 2}\D^1\D^2 D(t)
  &=
  ig^2
  \int \frac{\mathrm{d}^4k}{(2\pi)^4}
  \frac{
    ( k_1-\D_1/2)(k_2+\D_2/2)+ ( k_2-\D_2/2)(k_1+\D_1/2)
  }{
    \big[ (k-P)^2-m_2^2 \big]
    \big[ (k+\Delta/2)^2-m_1^2 \big]
    \big[ (k-\Delta/2)^2-m_1^2 \big]
  }
  \notag \\ &+
  ig^2
  \int \frac{\mathrm{d}^4k}{(2\pi)^4}
  \frac{
     ( k_1-\D_1/2)(k_2+\D_2/2)+ ( k_2-\D_2/2)(k_1+\D_1/2)
  }{
    \big[ (k-P)^2-m_1^2 \big]
    \big[ (k+\Delta/2)^2-m_2^2 \big]
    \big[ (k-\Delta/2)^2-m_2^2 \big]
  }
  \,.
\end{align}
In particular, we shall consider examples with masses appropriate
for a spin-less deuteron and a scalar pion.
In that case, we take $m_1=m_2=m$.
To simplify the notation, we additionally take $e_1+e_2=1$. 

The integrals above can be evaluated using Feynman parameters.
It is useful to start by considering the forward limit $\D^\mu=0$.
Then:
\begin{align}
  F(0)
  =
  {g^2\over 16\pi^2}\int_0^1\mathrm{d}x {(1-x)x\over \cM^2(x)}
  \,,
\end{align}
with
\begin{align}
  \cM^2(x)
  \equiv
  m^2-x(1-x)M^2
  \,.
\end{align}
The coupling constant $g$ is chosen to yield $F(0)=1$.
An important consistency check that
\begin{align}
  A(0)=F(0)
\end{align}
is satisfied.
We also find that
\begin{align}
  D(0) 
  =
  -{g^2\over 8\pi^2}\int_0^1\mathrm{d}x \,  (1- x){1-(1-x)^2/3\over \cM^2(x)}
  \label{D0}
  \,.
\end{align}
Defining the positive binding energy to be $B$ with $M=2m+B$ we find:
\begin{subequations}
  \begin{align}
    \lim_{B\to0}D(0)
    &=
    -{11\over3}+{32\over3\pi}\sqrt{B\over2M}-{\cal O}\left({B\over M}\right)
    \\
    \lim_{B\to \infty}D(0)
    &=
    -5+{2\over15}\left({2M\over M+B}\right)^2+{\mathcal O}\left({M^4\over B^4}\right)
    \,.
  \end{align}
\end{subequations}

An equivalent procedure~\cite{Miller:2009sg} to the use of Feynman parameters
is to use the Drell-Yan frame, where $\D^+=0$, and integrate over
$k ^-\equiv k^0-k^3$.
This enables one to obtain form factors in terms of light-front wave functions,
and also simplifies taking the non-relativistic limit.
We also use the relative momentum $\bfkappa= \bfk-x (\bfP-\bfDelta_\perp/2)$.  
The result is that:
\begin{align}
  F(t)
  =
  {1\over 2(2\pi)^3}
  \int \mathrm{d}^2\bfkappa
  \int_0^1 {\mathrm{d}x\over x(1-x)}
  \psi^*(x,\bfkappa+(1-x)\bfDelta_\perp)\psi(x,\bfkappa)
  \,,
  \label{2dft}
\end{align}
as found in Ref.~\cite{Gunion:1973ex},
with the frame-independent light-front wave function $\psi(x,\bfkappa)$ given by:
\begin{align}
  \psi(x,\bfkappa)
  \equiv
  g\left[M^2-{\bfkappa^2+m_1^2\over x}- {\bfkappa^2+m_2^2\over 1-x}\right]^{-1}
  \label{wf}
  \,,
\end{align}
where $\bfkappa$ is the $\perp$-component of the transverse relative momentum,
and $x$ is the fractional component of the longitudinal plus-component of the momentum
carried by the constituent of mass $m_1$.

For equal mass particles we find:
\begin{align}
  A(t)=F(t)
\end{align}
in this simple model,
and identity that is very useful for 
deuteron-like kinematics, as we shall see below.

To aid in the calculations that follow,
it is efficient to define a quantity:
\begin{align}
  G(\a,\bfDelta)
  &\equiv
  \int \mathrm{d}^2\bfkappa \,
  \psi^*(x,\bfkappa+\a\bfDelta_\perp)\psi(x,\bfkappa)
  \,,
\end{align}
with $\a$ as either $x$ or $1-x$,
depending on which particle is probed.
Then (for instance):
\begin{align}
  F(t)
  =
  {1\over 16\pi^3}\int_0^1{\mathrm{d}x\over x(1-x) }
  G(1-x,\bfDelta_\perp)
  \,.
\end{align}
The use of Feynman parameters leads to the result:
\begin{align}
  G(\a,\bfDelta)
  &=
  2\pi g^2 x^2{\bar x}^2\int_{0}^{1/2}\mathrm{d}z{1\over {\cal M}^2+\a^2Q^2(1/4-z^2)}
  =
  4\pi g^2 x^2{\bar x}^2\frac{
    \log \left(\frac{
      \sqrt{\alpha ^2 Q^2 \left(4 \cM^2+\alpha ^2 Q^2\right)}
      +
      2 \cM^2
      -
      \alpha^2 t
    }{2 \cM^2}\right)
 }{\sqrt{\alpha ^2 Q^2 \left(4 \cM^2+\alpha ^2 Q^2\right)}}
 \,.
\end{align}
Transverse densities are 2D Fourier transforms of the relevant form factors, such as:
\begin{align}
  \r_F(b_\perp)
  =
  \int {\mathrm{d}^2\bfDelta_\perp\over (2\pi)^2}
  e^{-i\bfDelta_\perp\cdot\bfb_\perp}F(t)
  \,.
\end{align}
so that it is useful to introduce the coordinate-space wave function:
\begin{align}
  \psi(x,\bfb_\perp)
  =
  \int {\mathrm{d}^2\kappa\over (2\pi)^2}
  \psi(x,\bfkappa)e^{i\bfkappa\cdot\bfb_\perp}
  \,.
\end{align}
In this model:
\begin{align}
  \psi(x,b_\perp)
  =
  {1\over (2\pi)^2}
  \int \mathrm{d}^2\bfkappa
  e^{i\bfkappa\cdot\bfb}\psi(x,\k)
  =
  {-g\over (2\pi)}x\bar x K_0(\cM b_\perp)
  \,,
\end{align}
where $K_0(x)$ is a modified Bessel function of the second kind,
and where $\bar x\equiv 1-x$.
In terms of this wave function, the density corresponding to the  electromagnetic form factor,, $ \rho_F$ is given by:
\begin{align}
  \r_F(b_\perp)
  =
  \int_0^1 {\mathrm{d}x \over (1-x)^2}
  \left|\psi\left(x,{b_\perp\over 1-x}\right)\right|^2
  \,.
\end{align}
Similarly, one may show that
\begin{align}
  \label{DDref}
  {1\over2}\D_1\D_2 D(t)
  =
  \int {\mathrm{d}^2\bfkappa\over (2\pi)^3}\int_0^1 {\mathrm{d}x\over x(1-x) }
  \Big(2\kappa_1\kappa_2-\D_1\D_2/2\Big)
  \left[
    {1\over x}\psi^*(x,\bfkappa+(1-x)\bfDelta_\perp/2)
    \,
    \psi(x,\bfkappa-(1-x)\bfDelta_\perp/2)
    \right]
\end{align}
Evaluation of the term proportional to $\kappa_1\kappa_2$
must result in a term proportional to $\D_1\D_2$.
However, it is worthwhile to find an explicit expression for $D(t)$.
This may be done by expressing $D(t)$ in terms of
the coordinate-space wave function, $\psi(x,\bfb_\perp)$.
Some algebra leads to the expression:
\begin{align}
  D(t)
  =
  -4\int_0^1 {\mathrm{d}x\over x(1-x) ^2}
  \int \mathrm{d}b_\perp \, b_\perp
  |\psi'(x,b_\perp)|^2 {J_2(x |\boldsymbol{\Delta}_\perp| b_\perp)  \over (-t)}
  -
  \int {\mathrm{d}x (1+x)\over x(1-x)}
  \int \mathrm{d}b_\perp \, b_\perp
  |\psi(x,b_\perp)|^2 J_0(x |\boldsymbol{\Delta}_\perp|b_\perp) 
  \,,\label{DT}
\end{align}
where:
$
   \psi'(x,b_\perp)
  \equiv
     { \partial\psi(x,b)\over \partial b}
  \,.
$
The quadrupole nature of $D(t)$
is exhibited by the appearance of the Bessel function of order 2 in \eq{DT}.

The transverse density is obtained in the same way.
For simplicity, we examine the light front version of
Polyakov's stress potential function $\widetilde{D}(b_\perp)$
defined in Eq.~(\ref{eqn:Dtilde}).
for which we find:
\begin{subequations}
  \begin{align}
    4P^+ \, \widetilde{D}(b_\perp)
    =
    {-1\over \pi}
    \int  {\mathrm{d}x\over x(1-x) }
    \left[
      f\left(x,\frac{b_\perp}{\barx}\right)
      +
      {1\over2}\bigg({1-x^2\over  x^2 \bar x}\,
      \left|\psi\left(x,\frac{b_\perp}{x}\right)\right|^2\bigg)
    \right]
  \,,
  \end{align}
  with 
  \begin{align}
    f(\a,b)
    \equiv
    \int_{b/\a}^\infty \mathrm{d}b'_\perp \, b'_\perp
    \left(1-{b^2\over \a^2{b'}^2}\right)
    |\psi'(\a,b')|^2.
  \end{align}
\end{subequations}
The above derivation of $D(t)$
exhibits the dependence on the wave function
and is thus useful in obtaining the non-relativistic limit.
One may instead proceed more directly to determine $D(t)$ by
starting with \eq{DDref} and using Feynman parameters to obtain:
\begin{align}
  D(t)
  =
  {-1\over 8 \pi^3}
  \int_0^1 \mathrm{d}x\,
  {1\over   x^2\bar x}
  \widetilde{G}(\bar x,t)
\end{align}
with
\begin{align}
  \widetilde{G}(\a,t)
  \equiv
  2 \pi g^2 x^2{\bar x}^2
  \int_{0}^{1}\mathrm{d}y\,
  {(1-y^2\a^2)\over {\cal M}^2-\a^2t/4(1-y^2)}
  =
  8 \pi g^2 x^2{\bar x}^2
  \,
  \frac{
    |\boldsymbol{\Delta}_\perp|
    -
    \frac{
      \left(4 - \left(\alpha ^2-1\right) t\right)
      \tanh^{-1}\left(\frac{\alpha  |\boldsymbol{\Delta}_\perp|}{\sqrt{4 - \alpha ^2 t} }\right)
    }{\alpha \sqrt{4-\alpha ^2 t}}
  }{
    |\boldsymbol{\Delta}_\perp|^3
  }
  \,.
\end{align}


\subsection{Breit-frame densities and transverse pseudo-densities}

The three dimensional Breit-frame densities
are given in Sec.~\ref{sec:densities:bf} above.
From these, one obtains the 2D Breit frame densities through
an Abel transform.
For spin-zero hadrons specifically,
one has:
\begin{align}
  \widetilde{D}_{\mathrm{BF}}(b_\perp)
  =
  \int_{-\infty}^\infty \mathrm{d}z \,
  \widetilde{D}_{\mathrm{BF}}(r)
  =
  \int{\mathrm{d}^2\boldsymbol{\Delta}_\perp \over (2\pi)^2}
  {D(t)\over \sqrt{1-t/(4M^2)}}
  e^{-i\bfDelta_\perp\cdot\bfb_\perp}
  \label{rdb}
  \,,
\end{align}
for example.
If $M\to\infty$ then
$\widetilde{D}_{\mathrm{BF}}(b_\perp) \rightarrow \widetilde{D}(b_\perp)$,
with the latter being the true (light front) relativistic density.
Since the integral goes over all values of  $t$, however,
the equality does not hold.
At best, one could have
$\widetilde{D}_{\mathrm{BF}}(b_\perp) \approx \widetilde{D}(b_\perp)$
if $D(-t\to4M^2)\approx0$.
Therefore we refer to
$\widetilde{D}_{\mathrm{BF}}(b_\perp)$
as a transverse \emph{pseudo-density}.

The integral appearing in \eq{rdb} provides a numerical challenge
because the asymptotic limit is $D(t)\sim \log^2(-t)/(-t)$.
This means that an expansion in powers of $-t^2/4M^2$ diverges.
A valid numerical procedure is obtained by
relating $\widetilde{D}_{\mathrm{BF}}(b_\perp)$
to the true transverse density $\widetilde{D}(b_\perp)$.
This is achieved by using Eq.~(\ref{eqn:Dtilde}) and the relation:
\begin{align}
  \int {\mathrm{d}^2\bfDelta_\perp\over (2\pi^2)}
  {e^{i\bfDelta_\perp\cdot\bfs}\over \sqrt{1-t^2/(4M^2)}}
  =
  {1\over \pi} {e^{-2Ms}\over s}
  \,,
\end{align}
so that:
\begin{align}
  \widetilde{D}_{\mathrm{BF}}(b_\perp)
  =
  {M\over \pi }\int \mathrm{d}^2s \,
  \widetilde{D}\big(|\bfb_\perp-{\bf s}|\big)
  {e^{-2Ms}\over s}
  \,.
\end{align}
This expression is amenable to two-dimensional  numerical integration.


\subsection{Non-relativistic limit}
 
The conventional lore is that the electromagnetic form factor is the three-dimensional
Fourier transform (3DFT) of the charge density.
This idea emerges only by taking the fully non-relativistic limit. 

We briefly review~\cite{Miller:2009sg} how the 3DFT emerges.
Our starting point is the wave function \eq{wf} and the form factor
\eq{2dft}.
Recall that the quantity
$x=k^+/P^+$. In the    non-relativistic
limit   the energy $k^0=m_1$, and $k^+=m_1+\kappa^3$, where
$\kappa^3$ is the third-component of  the relative longitudinal momentum.
Further we define the positive
binding energy $B$ so that 
\begin{align}
  M\equiv m_1+m_2-B
  \label{bdef}
  \,.
\end{align}
Then \cite{Brodsky:1989pv,Frankfurt:1981mk}:
\begin{align}
  x
  =
  \frac{m_1+\kappa^3}{M},\quad 1-x=\frac{M-m_1-\kappa^3}{M}=\frac{m_2-B-\kappa^3}{M}
  \label{a1}
  \,.
\end{align}
To obtain the non-relativistic wave function we express
the denominator appearing in \eq{wf} in terms of $\kappa^3$.
This gives:
\begin{align}
  M^2-{\kappa^2+m_1^2\over x}- {\kappa^2+m_2^2\over 1-x}
  \approx
  2M\left(-B-{\kappa^2\over 2\mu}\right)
  \label{prop12}
  \,,
\end{align}
where:
\begin{align}
  {\kappa}^2
  \equiv
  \bfkappa^2+\kappa_3^2
  \,,\qquad
  \boldsymbol{\kappa}
  =
  \bfkappa +\kappa^3 \hat{\bfz}
  \,, \qquad
  \mu\equiv{m_1m_2\over m_1+m_2}
  \,.
\end{align}
In deriving \eq{prop12},
we have dropped terms of order $(v/c)^3 = (k/m)^3$ and higher,
and terms of order $B/M$ and higher.
The result is that \eq{prop12} is
recognizable as $2M$ times the inverse of the non-relativistic propagator.

The next step is to determine the coordinate form of the
non-relativistic wave function $\psi_{\mathrm{NR}}(\mathbf{r})$ (where $\mathbf{r}$
is canonically conjugate to $\boldsymbol{\kappa}$)
and to show that the non-relativistic
form factor is a three-dimensional Fourier transform of 
$\left|\psi_{\mathrm{NR}}(\mathbf{r})\right|^2$.
First use the non-relativistic
approximation \eq{prop12} in \eq{wf} to find
\begin{align}
  \psi_{\mathrm{NR}}(\boldsymbol{\kappa})={-\mu g\over M(\kappa^2+\lambda^2)}
  \,, \qquad
  \lambda^2\equiv 2\mu B.
  \,.
  \label{nrp}
\end{align}
The coordinate-space wave function  $\psi_{\mathrm{NR}}(\mathbf{r})$ is given by
\begin{align}
  \psi_{\mathrm{NR}}(\mathbf{r})
  =
  {1\over (2\pi)^{3/2}}\int \mathrm{d}^3\boldsymbol{\kappa} \,
  e^{i\boldsymbol{\kappa}\cdot\mathbf{r}}
  \psi_{\mathrm{NR}}(\boldsymbol{\kappa})
  =
  -{\mu g\over 2M}\sqrt{\pi\over 2}{e^{-\lambda r}\over r}
  \label{nrr}
  \,.
\end{align}
The expression \eq{nrr} is seen as the standard result obtained for
the bound state of a two-particle system interacting via an attractive
delta function potential.
It is also the effective range approximation \cite{PhysRev.77.647},
now known as the leading-order term in effective field theory \cite{PhysRevC.59.617}.
 
The wave functions \eq{nrp} and \eq{nrr} 
enable us to examine the conditions needed for the approximations
\eq{a1} to be valid.
For \eq{a1} to work, we need $\kappa^2\ll m_{1,2}^2$.
The wave functions include all corrections to masses of order $\kappa^2/m_{1,2}^2$,
and therefore no further corrections
of order $\kappa/m_{1,2}$ or $B/m_{1,2}$ should be included.
Thus in evaluating the form factor we should use
\begin{subequations}
  \begin{align}
    \mathrm{d}x &\rightarrow {\mathrm{d}\boldsymbol{\kappa}^3\over M}
    \\
    x(1-x) &\rightarrow {\mu\over M }
    \\
    (1-x)\bfDelta &\rightarrow {m_2\over M}\bfDelta
    \,.
\label{rec}  \end{align}
\end{subequations}

The non-relativistic electromagnetic form factor
$F_{\mathrm{NR}}(t)$ is obtained by using
\eq{nrp} in the expression for the form factor \eq{2dft},
and taking the non-relativistic limit as defined above. 
The result is
\begin{align}
  F_{\mathrm{NR}}(t)
  =
  {1\over 2(2\pi)^3\mu}\int \mathrm{d}^3 \mathbf{r} \,
  \left|\psi_{\mathrm{NR}}(\mathbf{r})\right|^2
  e^{-i \bfDelta\cdot \bfr {m_2\over M}}
  \,.
\end{align}
This conforms to the commonplace expectation that the form factor is a 
three-dimensional Fourier transform of the density.
One may extract the density $|\psi_{\mathrm{NR}}(b_\perp)|^2$
by taking the Fourier transform of the form factor:
\begin{align}
  \int \mathrm{d}^3\boldsymbol{\Delta} \,
  F_{\mathrm{NR}}(t)
  e^{i\bfDelta\cdot\bfR}
  =
  {1\over 2\mu}\bigg|\psi_{\mathrm{NR}}\left({M\over m_2 }R\right)\bigg|^2
  \,.
\end{align}
Similarly, the non-relativistic gravitational form factor is given by
\begin{align}
  A_{\mathrm{NR}}(t)
  =
  {1\over 2(2\pi)^3\mu}
  \int \mathrm{d}^3 \mathbf{r} \,
  \left|\psi_{\mathrm{NR}}(\mathbf{r})\right|^2
  \bigg[
    {m_1} e^{-i \bfDelta\cdot \bfr {m_2\over m_1+m_2}}
    +
    {m_2}e^{i \bfDelta\cdot \bfr {m_1\over m_1+m_2}}
    \bigg]
  \,.
\end{align}
The principle difference between the relativistic and non-relativistic
computations of form factors occurs {\it e.g.} in \eq{rec}:
the variable factor $(1-x)$ is replaced by a constant.

We may evaluate the integrals immediately to find
\begin{subequations}
  \begin{align}
    F_{\mathrm{NR}}(t)
    &=
    {\tan^{-1}{|\boldsymbol{\Delta}_\perp| m_2\over 2(m_1+m_2)\lambda}
    \over
    {|\boldsymbol{\Delta}_\perp|m_2\over 2(m_1+m_2)\lambda}},\label{fnrgen}
    \\
    A_{\mathrm{NR}}(t)
    &=
    {2\lambda\over |\boldsymbol{\Delta}_\perp|}
    \bigg[
      {m_1\over m_2} \tan^{-1}\left({|\boldsymbol{\Delta}_\perp| m_2\over 2(m_1+m_2)\lambda}\right)
      +
      {m_2\over m_1} \tan^{-1}\left({|\boldsymbol{\Delta}_\perp| m_1\over 2(m_1+m_2)\lambda}\right)
      \bigg]
    \label{fnrgenA}
    \,,
  \end{align}
\end{subequations}
where the coupling constants and other constants enter in such a manner as
to make $F_{\mathrm{NR}}(0)=1$.
Note that if ${m_1=m_2}$ one has $A_{\mathrm{NR}}(t)=F_{\mathrm{NR}}(t)$,
as noted previously of the fully relativistic case.

It's easiest to get $D_{\mathrm{NR}}$ using momentum space techniques
First choose $g$ such that $F_{\mathrm{NR}}(0)=1$, which leads to 
\begin{align}
  g^2={16\pi M^2\l\over \mu}
  \,.
\end{align}
Then take the non-relativistic limit of \eq{DDref},
defining $\a\equiv m/M$.
We obtain:
\begin{subequations}
  \begin{align}
    \D_1\D_2
    D_{\mathrm{NR}}(t)
    &=
    \D_1\D_2
    {1\over 2\pi^3\mu}
    \int \mathrm{d}^3 \boldsymbol{\kappa}
    \Big(2 \kappa_1 \kappa_2-\D_1\D_2/2\Big)
    \psi_{\mathrm{NR}}^*(\boldsymbol \kappa+\a\boldsymbol\D/2)
    \psi_{\mathrm{NR}}(\boldsymbol\kappa-\a\boldsymbol\D/2)
    \\
    &=
    8\l/\pi^2
    \int \mathrm{d}^3 \boldsymbol{\kappa}
    \Big(2 \a^2z^2 \D_1\D_2-\D_1\D_2/2\Big)
    \int_{-1/2}^{1/2} \mathrm{d}z \,
    {1\over (\kappa^2+\l^2+\a^2\D^2(1/4-z^2))^2}
    \\
    D_{\mathrm{NR}}(t)
    &=
    -8\l
    \int_0^{1/2}\mathrm{d}z \,
    {(1-4\a^2z^2) \over \sqrt{\l^2-t (1/4-z^2)}}\label{eas}
    \\
    & =
    8\l \left(
    \frac{\lambda }{t}
    -
    \frac{
      \left(4 \lambda ^2-\left(\alpha ^2-2\right) t\right)
      \csc^{-1}\left(\sqrt{1-\frac{4 \lambda ^2}{\alpha ^2 t}}\right)
    }{
      2 \alpha  |\boldsymbol{\Delta}|^3
    }\right)
    \,.
  \end{align}
\end{subequations}
The value at zero momentum transfer is of interest,
and it can be obtained immediately from \eq{eas} to be 
\begin{align}
  D_{\mathrm{NR}}(0)
  =
  -4\left(1-{m^2\over 3M^2}\right)
  =
  -4\left(1-{1\over12}\left(1+{B\over M}\right)^2\right)
  \label{DNR0}
  \,.
\end{align}
For weak binding (with $m\approx M/2$),
we find $D_{\mathrm{NR}}(0)=-11/3$,
which agrees with the fully relativistic result in this limit.
For strong binding ($B\approx M$),
we find $D_{\mathrm{NR}}(0)=-8/3$.
For \emph{very strong} binding with $B/M>2\sqrt{3}-1\approx 2.5$,
we find $D_{\mathrm{NR}}(0)>0$, in violation of Polyakov's negativity condition.
This suggests that the non-relativistic model is invalid for such large binding energies.

To evaluate the non-relativistic transverse stress potential
$\widetilde{D}_{\mathrm{NR}}(b_\perp)$, we use the identity:
\begin{align}
  \int {\mathrm{d}^2\boldsymbol{\Delta}_\perp\over (2\pi)^2}
  e^{i\bfDelta_\perp\cdot\bfb_\perp}
  {1\over\sqrt{\l^2-t(1/4-z^2)}}
  =
  {1\over 2\pi b_\perp\sqrt{( 1/4-z^2)}}
  e^{- b_\perp\l\over\sqrt{1/4-z^2}},
\end{align}
giving us (with $\a=1/2$):
\begin{align}
  4 P^+ \,
  \widetilde{D}_{\mathrm{NR}}(b_\perp)
  =
  -{8\l\over 2\pi b_\perp}
  \int_0^{1/2}\mathrm{d}z\,
  {1-z^2\over\sqrt {1/4-z^2}}
  e^{- b_\perp\l\over\sqrt{1/4-z^2}}
  \label{rdnr}
  \,.
\end{align}


\subsection{Selected examples}

We examine two specific examples, with weak and strong binding respectively.
The former is appropriate for deuteron-like kinematics
and the latter for pion-like kinematics,
specially the kinematics in the light front model
of Ref.~\cite{Chung:1988mu}.


\subsubsection{Deuteron-like kinematics}

We first look at an example with weak binding,
namely deuteron-like kinematics with $M=1.875$~GeV
and $B=0.001 M$.
Recall that $M=2m+B$, with $B>0$.

\begin{figure}
  \includegraphics[width=0.45\textwidth]{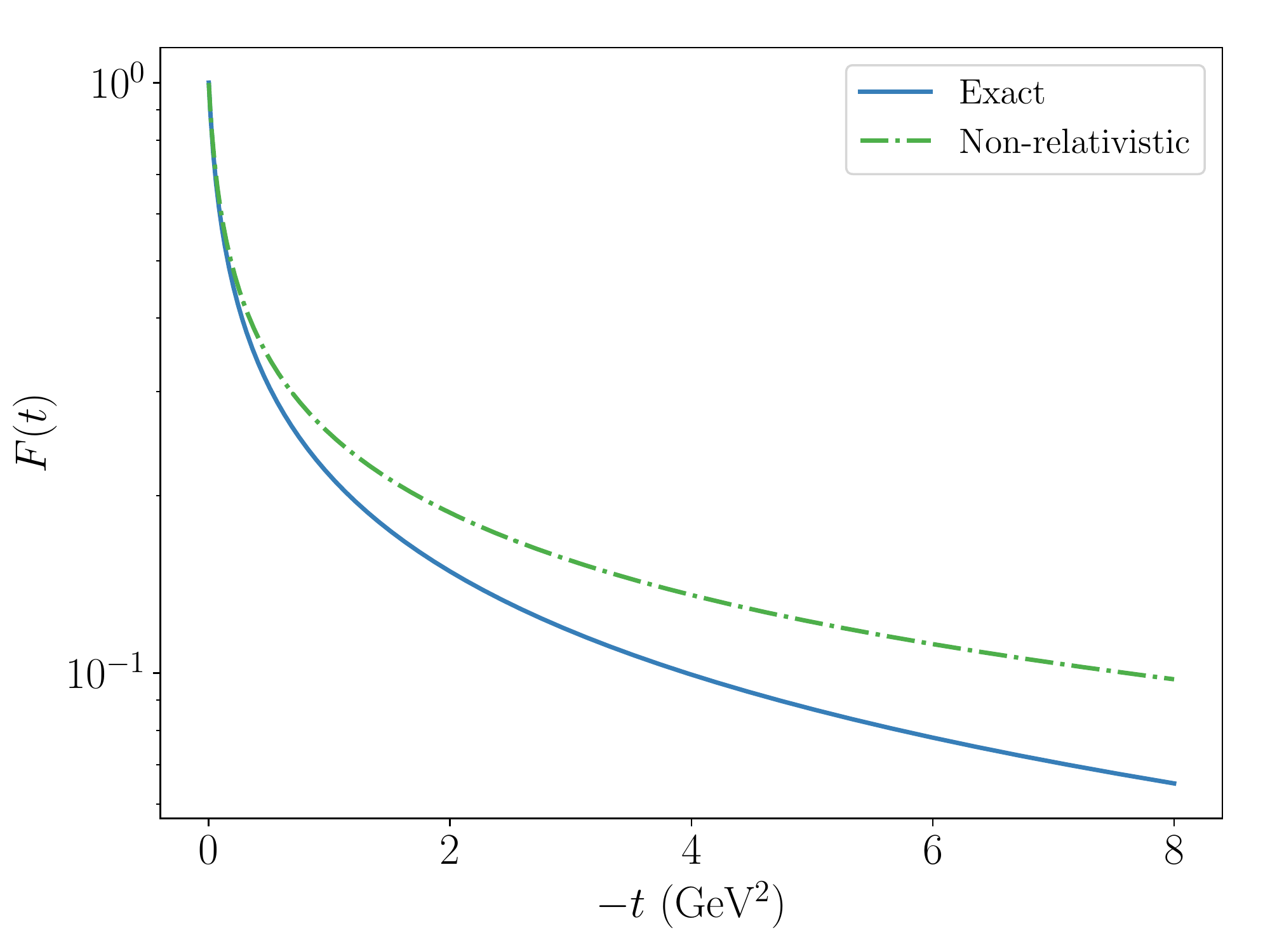} 
  \includegraphics[width=0.45\textwidth]{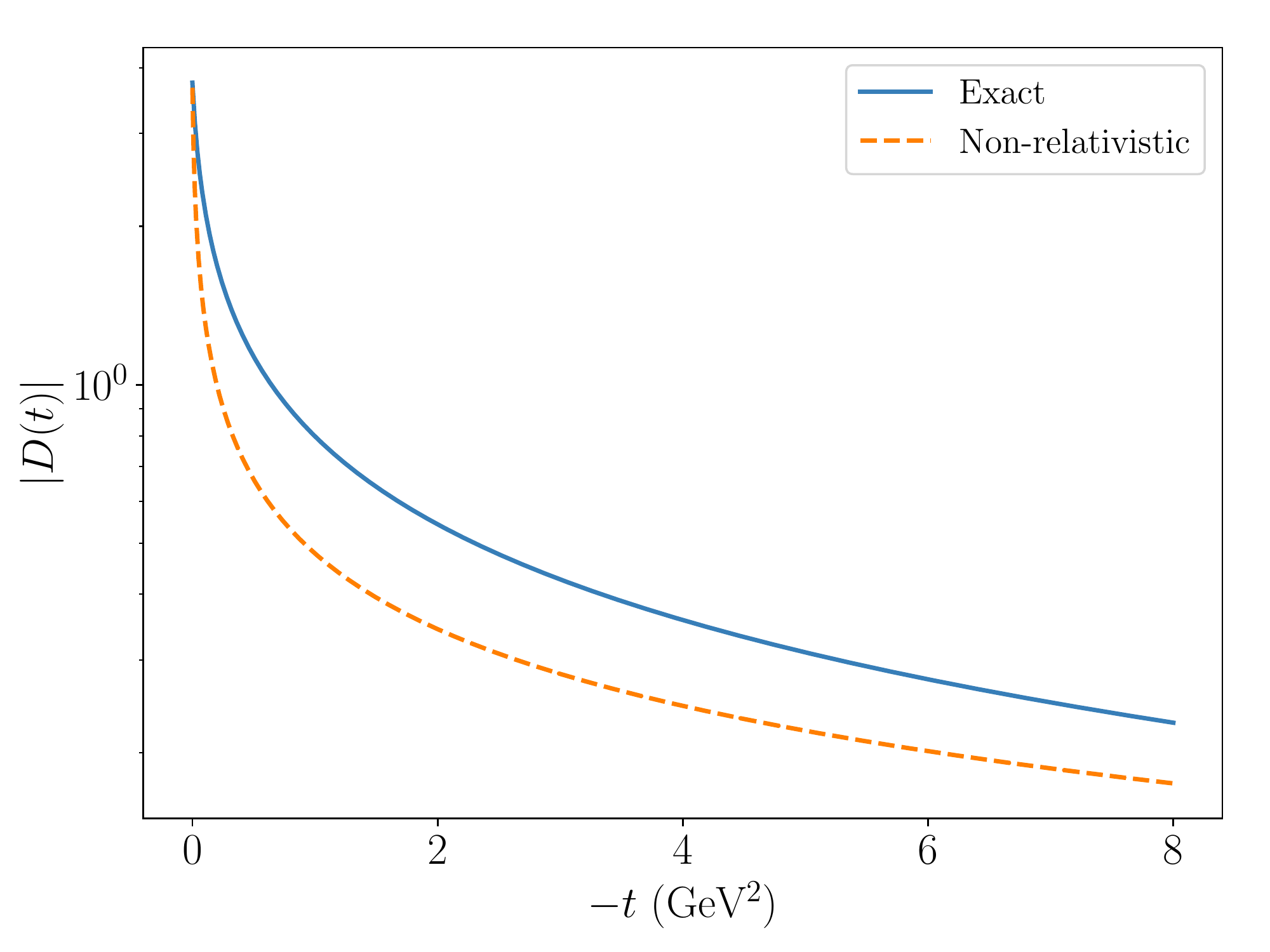} 
  \caption{
    Form factors for a scalar toy model with deuteron-like kinematics.
    Exact results for
    (left panel) $F(t)$ and (right panel) $D(t)$
    are compared to their non-relativistic approximations.
  }
  \label{fig:deuteronlike:FF}
\end{figure}

In Fig.~\ref{fig:deuteronlike:FF},
we compare the exact relativistic electromagnetic form factor $F(t)$
and exact stress form factor $D(t)$ to their
non-relativistic approximations
$F_{\mathrm{NR}}(t)$ and $D_{\mathrm{NR}}(t)$ as functions of $-t$.
The derivation of the exact and non-relativistic form factors
serves as a rough guide for the significance of relativistic effects in the system.

The exact and non-relativistic $F(t)$ are close when $-t/M^2 < 0.1$
(or $-t \approx 0.4$~GeV$^2$),
but diverge significantly at moderate and larger $-t$.
Since $D(0)$ is not protected by a conservation law
(unlike $F(0)$ or $A(0)$), it is possible for the exact and non-relativistic values to differ.
Indeed we find
an exact D-term value of $D(0)=-3.74$
and a non-relativistic value of $D_{\mathrm{NR}}(0)=-3.67$,
which are fairly close in magnitude,
but non-negligible.
The differences between the exact and non-relativistic form factors
illustrate the differences discussed in \eq{eqn:FNR:example} and related paragraphs.

\begin{figure}
  \includegraphics[width=0.45\textwidth]{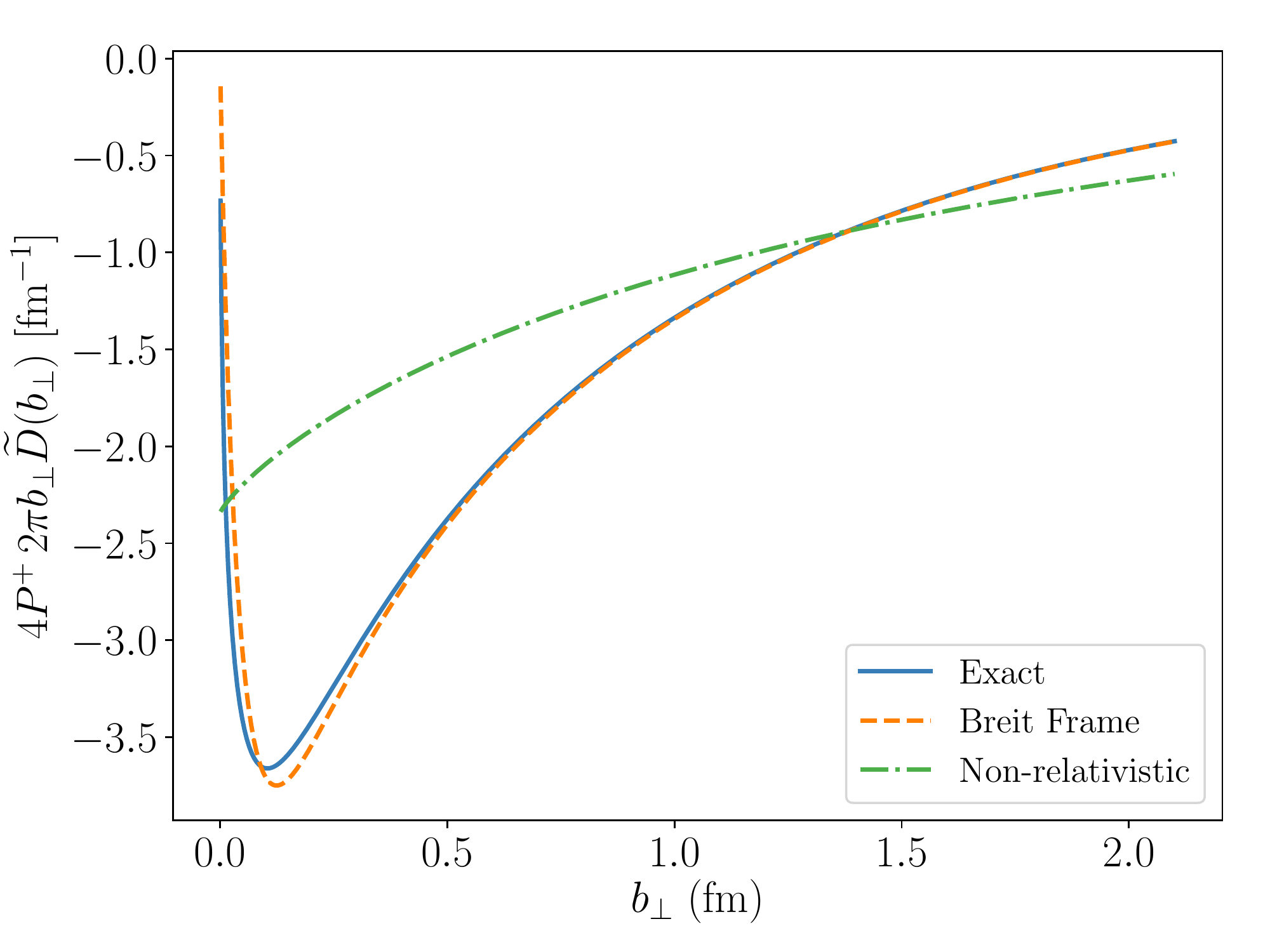} 
  \includegraphics[width=0.45\textwidth]{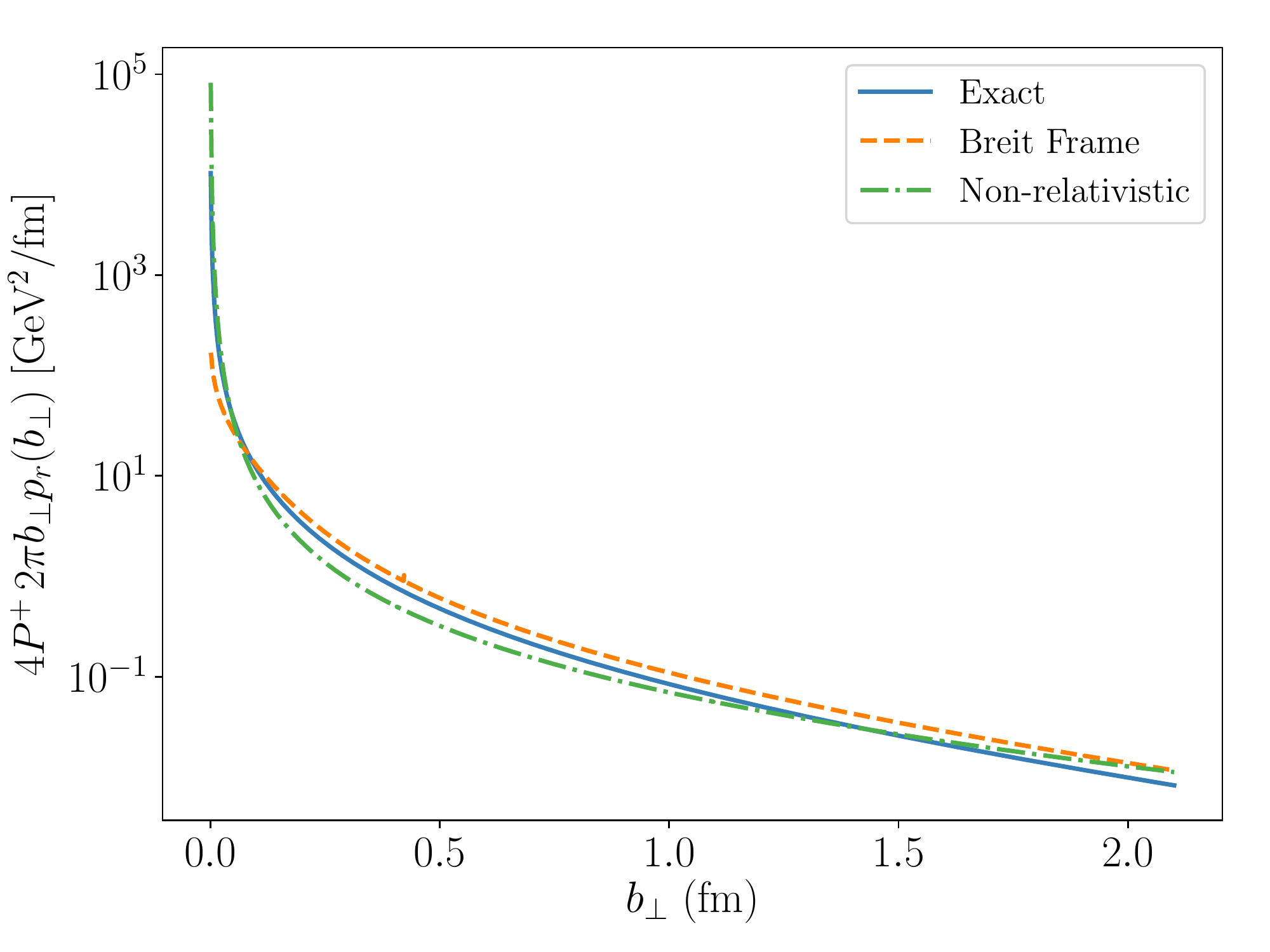} 
  \caption{
    $D$-term related densities
    for a scalar toy model with deuteron-like kinematics.
    Exact results, Breit frame results, and non-relativistic approximations for
    (left panel) the potential $\widetilde{D}(b_\perp)$ and
    (right panel) the 2D radial pressure $p_r(b_\perp)$
    are compared.
  }
  \label{fig:deuteronlike:density}
\end{figure}

Next, we compare the densities entailed by the relativistic and
non-relativistic $D(t)$ in Fig.~\ref{fig:deuteronlike:density}.
In particular, the direct Fourier transform $\widetilde{D}(b_\perp)$
[as defined in Eq.~(\ref{eqn:Dtilde})]
and the radial pressure are both examined.
For spin-zero targets, the Breit frame pseudo-density additionally differs from
both the exact light front result due to
the appearance of the factor $1/\sqrt{1-t/4M^2}$,
so it is compared to the exact and non-relativistic results in both cases.

In the left panel of Fig.~\ref{fig:deuteronlike:density},
the areas under the exact and Breit frame curves are the same,
since the two-dimensional integrals
of \eq{eqn:Dtilde} and \eq{rdb} are equal.
However, the factor $1/\sqrt{1-t/4M^2}$
in \eq{rdb} leads to the suppression of $\widetilde{D}_{\mathrm{BF}}(b_\perp)$
at small values of $b_\perp$ and enhancement at moderate values,
both by small amounts.
At large values, the densities become equal,
showing the corrections related to the factor $1/\sqrt{1-t/4M^2}$
become negligible at large distances.
However,
the non-relativistic result for $\widetilde{D}(b_\perp)$
is \emph{significantly} different from the exact and Breit frame results,
suggesting that relativistic effects may persist in mechanical densities
even at fairly large distances.

This same trend can be seen in the radial pressure,
as depicted in 
the right panel of Fig.~\ref{fig:deuteronlike:density}.
Relativistic effects propagate to large $b_\perp$,
even when they would be expected to vanish.
This occurs because the pressure does not correspond to a conserved current,
and is thus sensitive to the details the dynamics of a system
(see Refs.~\cite{Hudson:2017oul,Freese:2019bhb} for
examples of cases where the details of dynamics are significant).
The details of the dynamics affect the overall mechanical structure of the hadron,
and not just local aspects of the structure at small distances.
Note that the log scale in the right panel of Fig.~\ref{fig:deuteronlike:density}
covers more than six orders of magnitude,
so that the apparently small differences are actually rather large.

For a slightly more detailed perspective,
consider the right hand side of Eq.~(\ref{DDref}),
which was used in obtaining the relativistic $\widetilde{D}(b_\perp)$.
The matrix element is weighted by a factor $\frac{1}{x}$
compared to the matrix elements for the densities associated with $F(t)$ and $A(t)$.
This factor increases the integrand when $x\sim0$ or  $x\sim1$,
conditions that are explicitly discounted by the non-relativistic limit
in which (for equal mass constituents)
we make the replacement $\frac{1}{x} \mapsto \frac{1}{\alpha} \approx 2$.

The overall lesson of this case is that,
because the mechanical properties of a hadron are not protected by
a conservation law,
they are sensitive to the details of dynamics,
and accordingly non-relativistic effects can have a significant effect
even at large distances and even for weakly bound systems.


\subsubsection{Pion-like kinematics}

We now consider the scalar toy model with kinematics appropriate
for a constituent quark model of the pion.
We use a ``quark'' mass of 210~MeV,
which leads to an excellent description of the pion's
electromagnetic form factor~\cite{Chung:1988mu}.
With $M = 140~MeV$, this means that $B/M=2$,
and according to \eq{DNR0} we get $D_{\mathrm{NR}}(0)=-1$,
a result that immediately demonstrates the importance of relativistic effects for this model. 

\begin{figure}
  \includegraphics[width=0.45\textwidth]{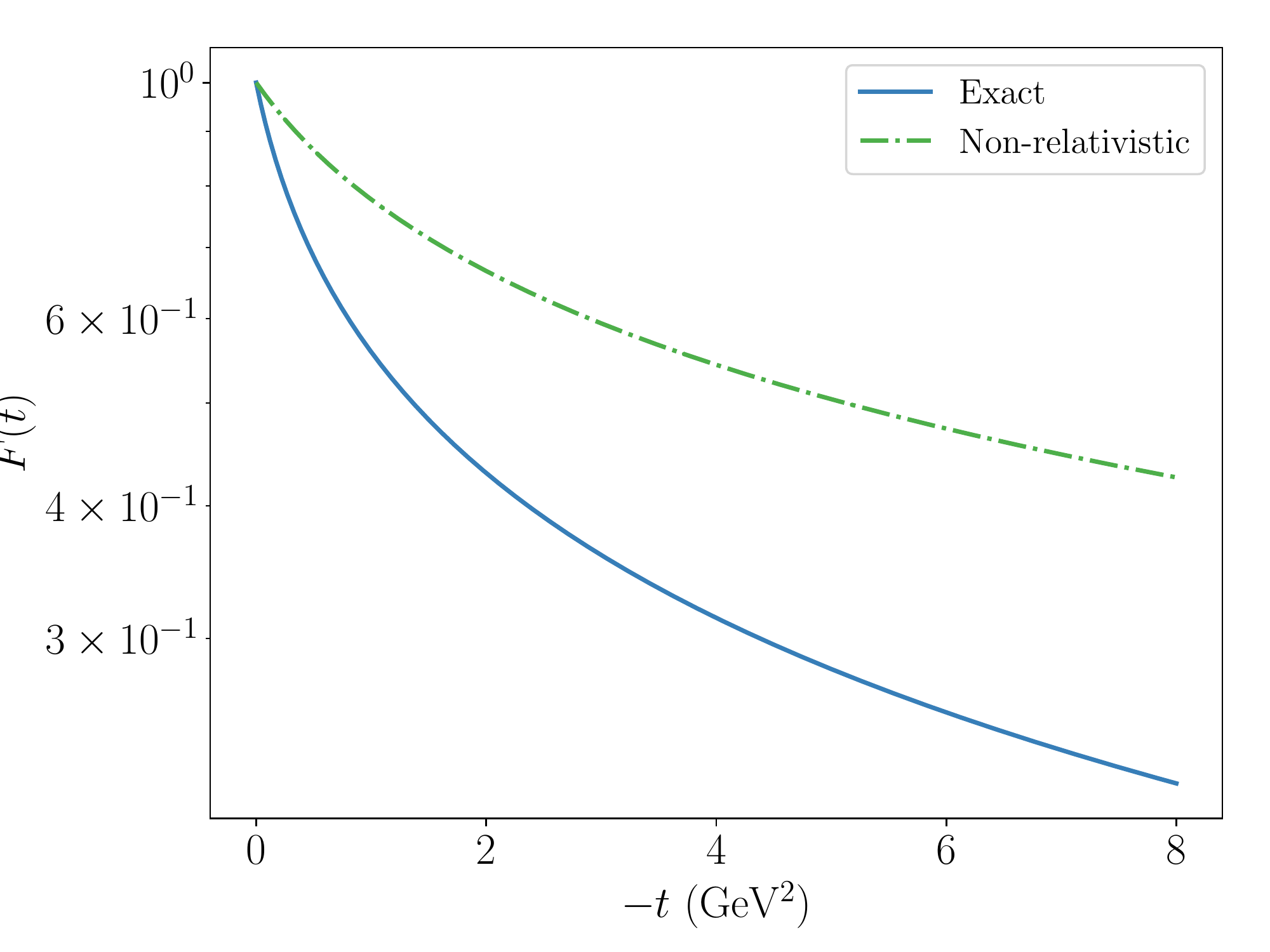} 
  \includegraphics[width=0.45\textwidth]{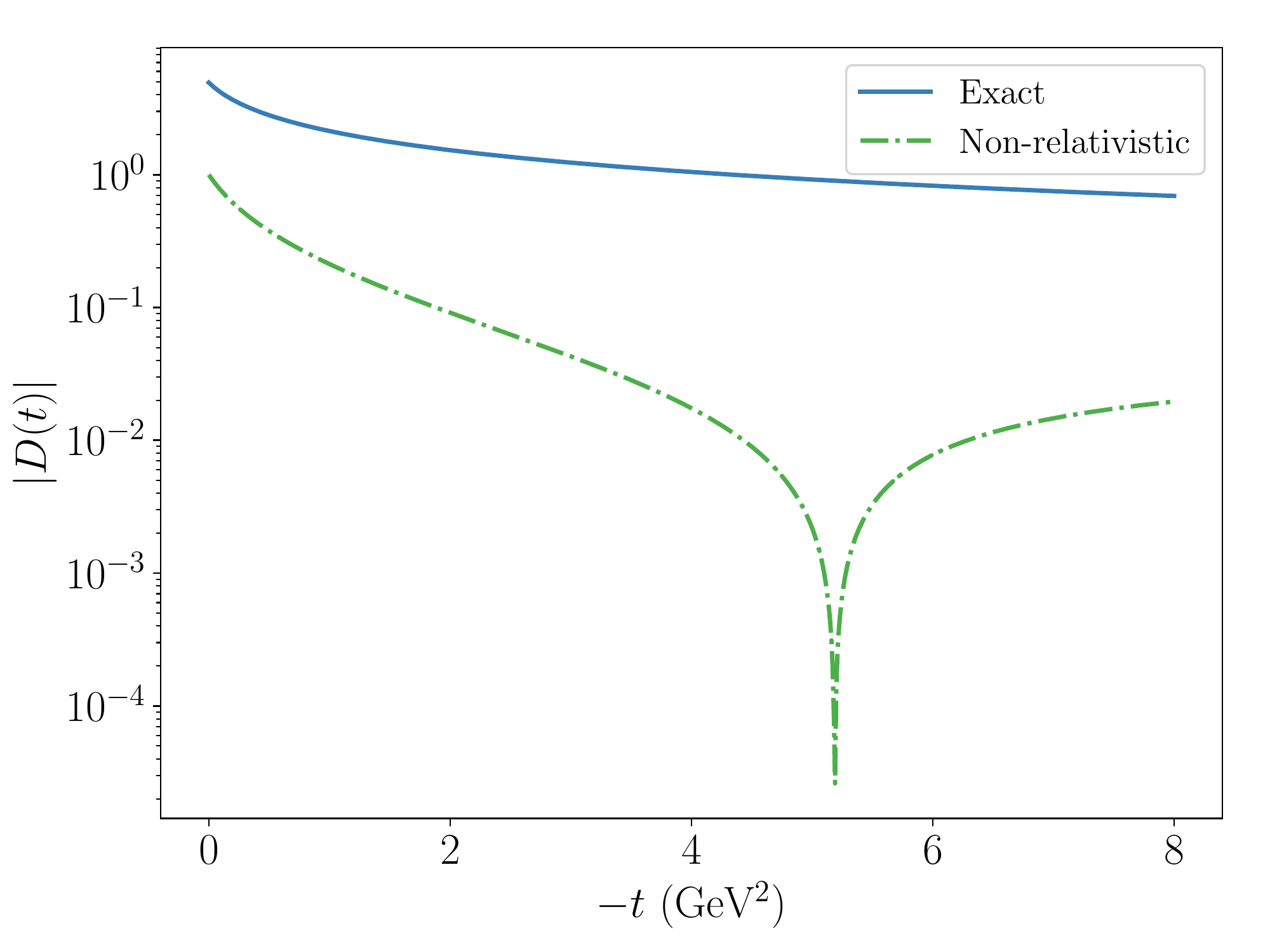} 
  \caption{
    Form factors for a scalar toy model with pion-like kinematics.
    Exact results for
    (left panel) $F(t)$ and (right panel) $D(t)$
    are compared to their non-relativistic approximations.
  }
  \label{fig:pionlike:FF}
\end{figure}

In Fig.~\ref{fig:pionlike:FF}, we present the exact and non-relativistic
form factors $F(t)$ and $D(t)$.
As expected, the relativistic effects are substantial.
Even at $t=0$, we have $D(0) = -4.9$ and $D_{\mathrm{NR}}(0) = -1$,
about a fifth of the true value.
Moreover, the non-relativistic approximation of $D(t)$ as a zero crossing
that's absent in the exact result,
as seen in the right panel of Fig.~\ref{fig:pionlike:FF}.
This is because of the increasing importance of
$z$ values near $\frac{1}{2}$ with larger $\D^2$,
which causes the second term in the integrand of \eq{eas}
to dominate over the first term.

\begin{figure}
  \includegraphics[width=0.45\textwidth]{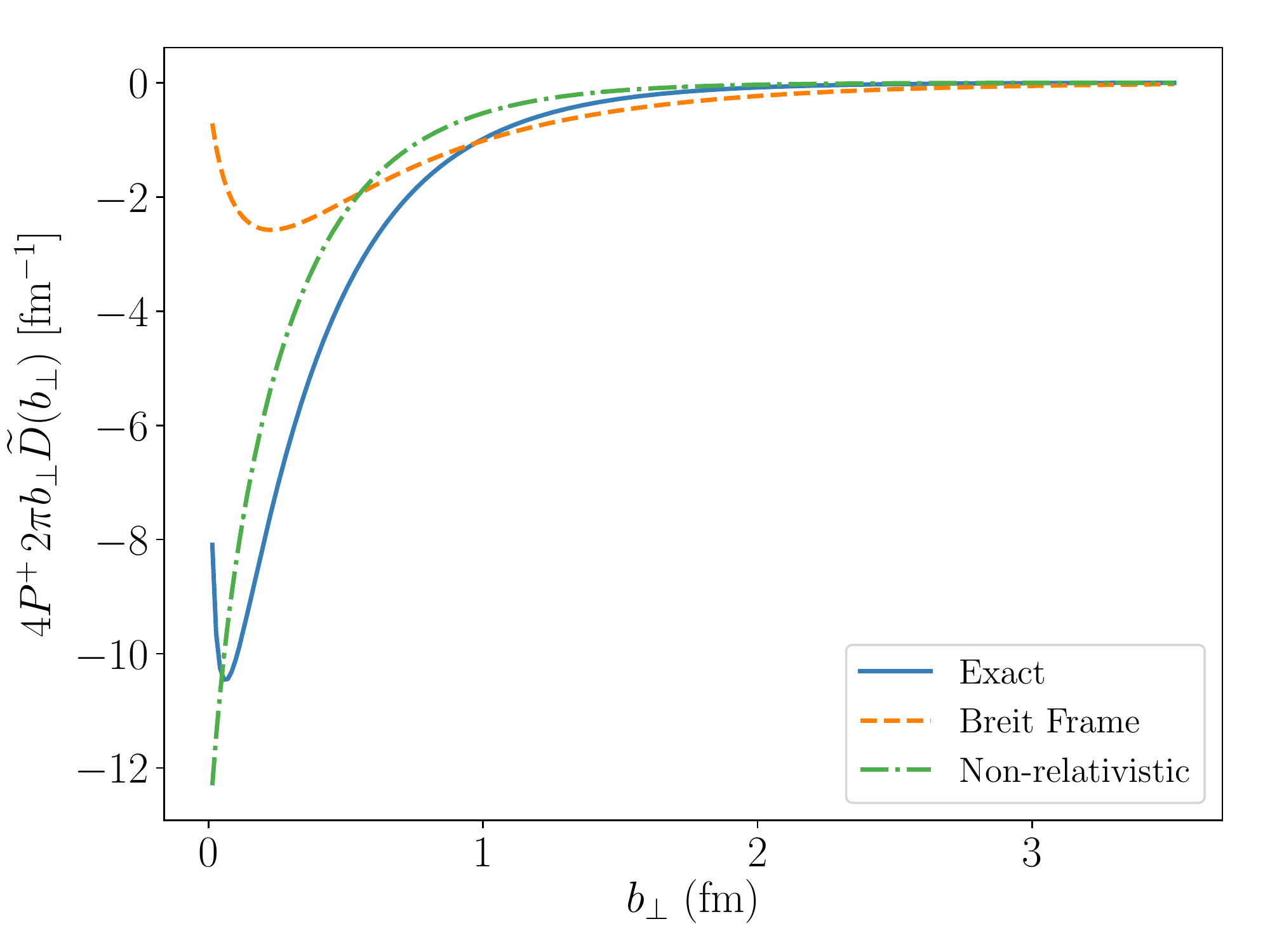} 
  \includegraphics[width=0.45\textwidth]{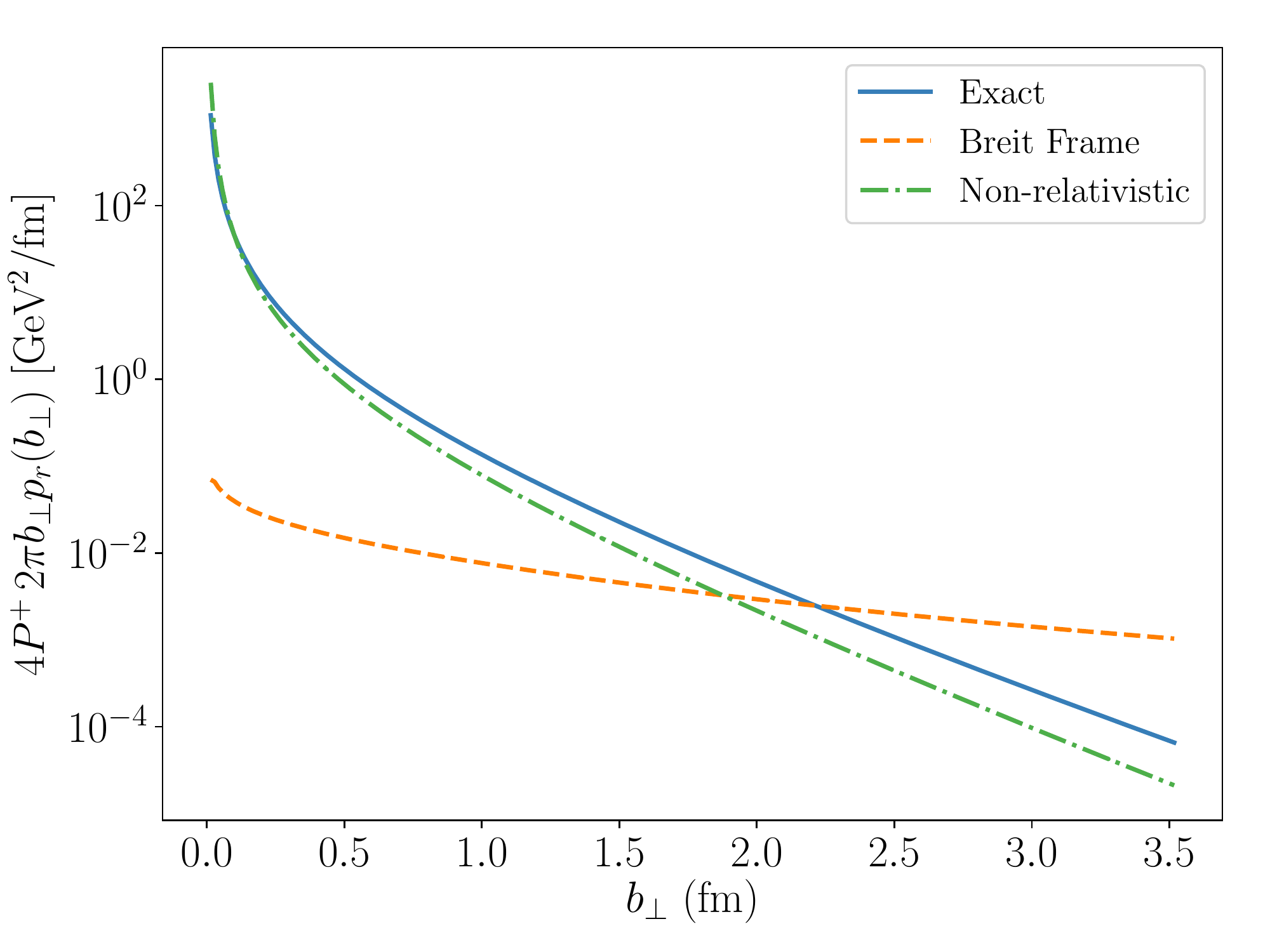} 
  \caption{
    $D$-term related densities
    for a scalar toy model with pion-like kinematics.
    Exact results, Breit frame results, and non-relativistic approximations for
    (left panel) the potential $\widetilde{D}(b_\perp)$ and
    (right panel) the 2D radial pressure $p_r(b_\perp)$
    are compared.
  }
  \label{fig:pionlike:density}
\end{figure}

The density $\widetilde{D}(b_\perp)$ and radial pressure
are shown in Fig.~\ref{fig:pionlike:density}.
As expected, there are very substantial differences between
the true density, Breit frame pseudo-density, and non-relativistic approximation.
Remarkably, the Breit frame result is a worse approximation to the true
density in this case than the non-relativistic approximation.
This demonstrates the significance of the extraneous factor
$1/\sqrt{1-t/(4M^2)}$ present in the Breit frame densities,
and strongly forces us to the conclusion that the findings of
Refs.~\cite{Panteleeva:2021iip,Kim:2021jjf}
cannot be applied outside of the spin-half case,
where those results hold only by accident.


\section{Summary and conclusions}
\label{sec:conclusion}

In this work, we obtained exact relativistic (light front)
and approximate non-relativistic expressions
for densities in spin-zero in spin-half hadrons.
Focus was placed on the $P^+$ and mass densities,
as well as the pressures encoded by the stress tensor as seen from
the perspective of an observer comoving with the hadron.
We compared the exact and non-relativistic expressions to those
obtained in the Breit frame formalism.
We find that, in general, the Breit frame densities do not have
a direct correspondence with the exact light front densities,
even through Abel transforms.
This failure of correspondence occurs
in the spin-zero case because of an extraneous factor
$1/\sqrt{1-t/(4M^2)}$ present in the integrand of every spin-zero
Breit frame density,
and in both cases because the relativistic light front densities
do not exhibit spherical symmetry.
The latter of these facts is illustrated for spin-half hadrons in particular
by the azimuthal dependence of $P^+$ densities
and pressures of transversely polarized states,
to which the inverse Abel transform is inapplicable even formally.
In general, however, the light front formalism lacks spherical
symmetry, since there is no $\mathrm{SO}(3)$ subgroup
of the Poincar\'e group that commutes with $P^-$~\cite{Brodsky:1997de}.

The significance of both relativistic effects
and the extraneous term
$1/\sqrt{1-t/(4M^2)}$ present in the spin-zero Breit frame densities
is illustrated through a pedagogical model.
Relativistic effects were found to affect the mechanical structure of
a composite system significantly, even for weakly bound systems
and at large distances---in contrast to the electromagnetic density
or mass density, both of which are protected by conservation laws.
Moreover, for strongly bound systems,
we found that the extraneous factor
$1/\sqrt{1-t/(4M^2)}$ makes the (Abel transform of the)
Breit frame density a poor approximation to the true density.

Taking the inverse Abel transform of a transverse light front density can,
at best, return a partially non-relativistic approximation.
This approximation is partially non-relativistic,
since a non-relativistic approximation of the internal dynamics has not been
applied to the form factors associated with the density,
but instead only to the accompanying Lorentz tensors.
In some circumstances, such as neutron star structure,
this approximation \emph{may} be warranted
(cf.\ Ref.~\cite{Rajan:2018zzy} for an example of this application),
but one should bear in mind that this operation is an approximation
that eliminates effects due to boosts from the target's rest frame.
For targets with wave functions localized to a smaller distance than their
reduced Compton wavelength, or targets for which a finer resolution
of internal structure than the Compton wavelength is
desired---such as hadrons---this approximation
cannot be justified.

\begin{acknowledgments}
  We would like to thank
  Matthias Burkardt, Wim Cosyn, Xiangdong Ji, and Simonetta Liuti
  for illuminating discussions on the topics covered in this paper.
  This work was supported by the U.S.\ Department of Energy
  Office of Science, Office of Nuclear Physics under Award Number
  DE-FG02-97ER-41014.
\end{acknowledgments}


\bibliography{main.bib}

\begin{thebibliography}{45}%
\makeatletter
\providecommand \@ifxundefined [1]{%
 \@ifx{#1\undefined}
}%
\providecommand \@ifnum [1]{%
 \ifnum #1\expandafter \@firstoftwo
 \else \expandafter \@secondoftwo
 \fi
}%
\providecommand \@ifx [1]{%
 \ifx #1\expandafter \@firstoftwo
 \else \expandafter \@secondoftwo
 \fi
}%
\providecommand \natexlab [1]{#1}%
\providecommand \enquote  [1]{``#1''}%
\providecommand \bibnamefont  [1]{#1}%
\providecommand \bibfnamefont [1]{#1}%
\providecommand \citenamefont [1]{#1}%
\providecommand \href@noop [0]{\@secondoftwo}%
\providecommand \href [0]{\begingroup \@sanitize@url \@href}%
\providecommand \@href[1]{\@@startlink{#1}\@@href}%
\providecommand \@@href[1]{\endgroup#1\@@endlink}%
\providecommand \@sanitize@url [0]{\catcode `\\12\catcode `\$12\catcode
  `\&12\catcode `\#12\catcode `\^12\catcode `\_12\catcode `\%12\relax}%
\providecommand \@@startlink[1]{}%
\providecommand \@@endlink[0]{}%
\providecommand \url  [0]{\begingroup\@sanitize@url \@url }%
\providecommand \@url [1]{\endgroup\@href {#1}{\urlprefix }}%
\providecommand \urlprefix  [0]{URL }%
\providecommand \Eprint [0]{\href }%
\providecommand \doibase [0]{http://dx.doi.org/}%
\providecommand \selectlanguage [0]{\@gobble}%
\providecommand \bibinfo  [0]{\@secondoftwo}%
\providecommand \bibfield  [0]{\@secondoftwo}%
\providecommand \translation [1]{[#1]}%
\providecommand \BibitemOpen [0]{}%
\providecommand \bibitemStop [0]{}%
\providecommand \bibitemNoStop [0]{.\EOS\space}%
\providecommand \EOS [0]{\spacefactor3000\relax}%
\providecommand \BibitemShut  [1]{\csname bibitem#1\endcsname}%
\let\auto@bib@innerbib\@empty
\bibitem [{\citenamefont {Kobzarev}\ and\ \citenamefont
  {Okun}(1962)}]{Kobzarev:1962wt}%
  \BibitemOpen
  \bibfield  {author} {\bibinfo {author} {\bibfnamefont {I.~Y.}\ \bibnamefont
  {Kobzarev}}\ and\ \bibinfo {author} {\bibfnamefont {L.~B.}\ \bibnamefont
  {Okun}},\ }\href@noop {} {\bibfield  {journal} {\bibinfo  {journal} {Zh.
  Eksp. Teor. Fiz.}\ }\textbf {\bibinfo {volume} {43}},\ \bibinfo {pages}
  {1904} (\bibinfo {year} {1962})}\BibitemShut {NoStop}%
\bibitem [{\citenamefont {Ji}(1995{\natexlab{a}})}]{Ji:1994av}%
  \BibitemOpen
  \bibfield  {author} {\bibinfo {author} {\bibfnamefont {X.-D.}\ \bibnamefont
  {Ji}},\ }\href {\doibase 10.1103/PhysRevLett.74.1071} {\bibfield  {journal}
  {\bibinfo  {journal} {Phys. Rev. Lett.}\ }\textbf {\bibinfo {volume} {74}},\
  \bibinfo {pages} {1071} (\bibinfo {year} {1995}{\natexlab{a}})},\ \Eprint
  {http://arxiv.org/abs/hep-ph/9410274} {arXiv:hep-ph/9410274} \BibitemShut
  {NoStop}%
\bibitem [{\citenamefont {Ji}(1995{\natexlab{b}})}]{Ji:1995sv}%
  \BibitemOpen
  \bibfield  {author} {\bibinfo {author} {\bibfnamefont {X.-D.}\ \bibnamefont
  {Ji}},\ }\href {\doibase 10.1103/PhysRevD.52.271} {\bibfield  {journal}
  {\bibinfo  {journal} {Phys. Rev. D}\ }\textbf {\bibinfo {volume} {52}},\
  \bibinfo {pages} {271} (\bibinfo {year} {1995}{\natexlab{b}})},\ \Eprint
  {http://arxiv.org/abs/hep-ph/9502213} {arXiv:hep-ph/9502213} \BibitemShut
  {NoStop}%
\bibitem [{\citenamefont {Lorc\'e}(2018)}]{Lorce:2017xzd}%
  \BibitemOpen
  \bibfield  {author} {\bibinfo {author} {\bibfnamefont {C.}~\bibnamefont
  {Lorc\'e}},\ }\href {\doibase 10.1140/epjc/s10052-018-5561-2} {\bibfield
  {journal} {\bibinfo  {journal} {Eur. Phys. J. C}\ }\textbf {\bibinfo {volume}
  {78}},\ \bibinfo {pages} {120} (\bibinfo {year} {2018})},\ \Eprint
  {http://arxiv.org/abs/1706.05853} {arXiv:1706.05853 [hep-ph]} \BibitemShut
  {NoStop}%
\bibitem [{\citenamefont {Hatta}\ \emph {et~al.}(2018)\citenamefont {Hatta},
  \citenamefont {Rajan},\ and\ \citenamefont {Tanaka}}]{Hatta:2018sqd}%
  \BibitemOpen
  \bibfield  {author} {\bibinfo {author} {\bibfnamefont {Y.}~\bibnamefont
  {Hatta}}, \bibinfo {author} {\bibfnamefont {A.}~\bibnamefont {Rajan}}, \ and\
  \bibinfo {author} {\bibfnamefont {K.}~\bibnamefont {Tanaka}},\ }\href
  {\doibase 10.1007/JHEP12(2018)008} {\bibfield  {journal} {\bibinfo  {journal}
  {JHEP}\ }\textbf {\bibinfo {volume} {12}},\ \bibinfo {pages} {008} (\bibinfo
  {year} {2018})},\ \Eprint {http://arxiv.org/abs/1810.05116} {arXiv:1810.05116
  [hep-ph]} \BibitemShut {NoStop}%
\bibitem [{\citenamefont {Ashman}\ \emph {et~al.}(1988)\citenamefont {Ashman}
  \emph {et~al.}}]{Ashman:1987hv}%
  \BibitemOpen
  \bibfield  {author} {\bibinfo {author} {\bibfnamefont {J.}~\bibnamefont
  {Ashman}} \emph {et~al.} (\bibinfo {collaboration} {European Muon}),\ }\href
  {\doibase 10.1016/0370-2693(88)91523-7} {\bibfield  {journal} {\bibinfo
  {journal} {Phys. Lett. B}\ }\textbf {\bibinfo {volume} {206}},\ \bibinfo
  {pages} {364} (\bibinfo {year} {1988})}\BibitemShut {NoStop}%
\bibitem [{\citenamefont {Ji}(1997)}]{Ji:1996ek}%
  \BibitemOpen
  \bibfield  {author} {\bibinfo {author} {\bibfnamefont {X.-D.}\ \bibnamefont
  {Ji}},\ }\href {\doibase 10.1103/PhysRevLett.78.610} {\bibfield  {journal}
  {\bibinfo  {journal} {Phys. Rev. Lett.}\ }\textbf {\bibinfo {volume} {78}},\
  \bibinfo {pages} {610} (\bibinfo {year} {1997})},\ \Eprint
  {http://arxiv.org/abs/hep-ph/9603249} {arXiv:hep-ph/9603249} \BibitemShut
  {NoStop}%
\bibitem [{\citenamefont {Leader}\ and\ \citenamefont
  {Lorc\'e}(2014)}]{Leader:2013jra}%
  \BibitemOpen
  \bibfield  {author} {\bibinfo {author} {\bibfnamefont {E.}~\bibnamefont
  {Leader}}\ and\ \bibinfo {author} {\bibfnamefont {C.}~\bibnamefont
  {Lorc\'e}},\ }\href {\doibase 10.1016/j.physrep.2014.02.010} {\bibfield
  {journal} {\bibinfo  {journal} {Phys. Rept.}\ }\textbf {\bibinfo {volume}
  {541}},\ \bibinfo {pages} {163} (\bibinfo {year} {2014})},\ \Eprint
  {http://arxiv.org/abs/1309.4235} {arXiv:1309.4235 [hep-ph]} \BibitemShut
  {NoStop}%
\bibitem [{\citenamefont {Polyakov}(2003)}]{Polyakov:2002yz}%
  \BibitemOpen
  \bibfield  {author} {\bibinfo {author} {\bibfnamefont {M.~V.}\ \bibnamefont
  {Polyakov}},\ }\href {\doibase 10.1016/S0370-2693(03)00036-4} {\bibfield
  {journal} {\bibinfo  {journal} {Phys. Lett. B}\ }\textbf {\bibinfo {volume}
  {555}},\ \bibinfo {pages} {57} (\bibinfo {year} {2003})},\ \Eprint
  {http://arxiv.org/abs/hep-ph/0210165} {arXiv:hep-ph/0210165} \BibitemShut
  {NoStop}%
\bibitem [{\citenamefont {Polyakov}\ and\ \citenamefont
  {Schweitzer}(2018)}]{Polyakov:2018zvc}%
  \BibitemOpen
  \bibfield  {author} {\bibinfo {author} {\bibfnamefont {M.~V.}\ \bibnamefont
  {Polyakov}}\ and\ \bibinfo {author} {\bibfnamefont {P.}~\bibnamefont
  {Schweitzer}},\ }\href {\doibase 10.1142/S0217751X18300259} {\bibfield
  {journal} {\bibinfo  {journal} {Int. J. Mod. Phys. A}\ }\textbf {\bibinfo
  {volume} {33}},\ \bibinfo {pages} {1830025} (\bibinfo {year} {2018})},\
  \Eprint {http://arxiv.org/abs/1805.06596} {arXiv:1805.06596 [hep-ph]}
  \BibitemShut {NoStop}%
\bibitem [{\citenamefont {Lorc\'e}\ \emph {et~al.}(2019)\citenamefont
  {Lorc\'e}, \citenamefont {Moutarde},\ and\ \citenamefont
  {Trawi\'nski}}]{Lorce:2018egm}%
  \BibitemOpen
  \bibfield  {author} {\bibinfo {author} {\bibfnamefont {C.}~\bibnamefont
  {Lorc\'e}}, \bibinfo {author} {\bibfnamefont {H.}~\bibnamefont {Moutarde}}, \
  and\ \bibinfo {author} {\bibfnamefont {A.~P.}\ \bibnamefont {Trawi\'nski}},\
  }\href {\doibase 10.1140/epjc/s10052-019-6572-3} {\bibfield  {journal}
  {\bibinfo  {journal} {Eur. Phys. J. C}\ }\textbf {\bibinfo {volume} {79}},\
  \bibinfo {pages} {89} (\bibinfo {year} {2019})},\ \Eprint
  {http://arxiv.org/abs/1810.09837} {arXiv:1810.09837 [hep-ph]} \BibitemShut
  {NoStop}%
\bibitem [{\citenamefont {Freese}\ and\ \citenamefont
  {Miller}(2021{\natexlab{a}})}]{Freese:2021czn}%
  \BibitemOpen
  \bibfield  {author} {\bibinfo {author} {\bibfnamefont {A.}~\bibnamefont
  {Freese}}\ and\ \bibinfo {author} {\bibfnamefont {G.~A.}\ \bibnamefont
  {Miller}},\ }\href@noop {} {\  (\bibinfo {year} {2021}{\natexlab{a}})},\
  \Eprint {http://arxiv.org/abs/2102.01683} {arXiv:2102.01683 [hep-ph]}
  \BibitemShut {NoStop}%
\bibitem [{\citenamefont {Burkert}\ \emph {et~al.}(2018)\citenamefont
  {Burkert}, \citenamefont {Elouadrhiri},\ and\ \citenamefont
  {Girod}}]{Burkert:2018bqq}%
  \BibitemOpen
  \bibfield  {author} {\bibinfo {author} {\bibfnamefont {V.~D.}\ \bibnamefont
  {Burkert}}, \bibinfo {author} {\bibfnamefont {L.}~\bibnamefont
  {Elouadrhiri}}, \ and\ \bibinfo {author} {\bibfnamefont {F.~X.}\ \bibnamefont
  {Girod}},\ }\href {\doibase 10.1038/s41586-018-0060-z} {\bibfield  {journal}
  {\bibinfo  {journal} {Nature}\ }\textbf {\bibinfo {volume} {557}},\ \bibinfo
  {pages} {396} (\bibinfo {year} {2018})}\BibitemShut {NoStop}%
\bibitem [{\citenamefont {Dutrieux}\ \emph {et~al.}(2021)\citenamefont
  {Dutrieux}, \citenamefont {Lorc\'e}, \citenamefont {Moutarde}, \citenamefont
  {Sznajder}, \citenamefont {Trawi\'nski},\ and\ \citenamefont
  {Wagner}}]{Dutrieux:2021nlz}%
  \BibitemOpen
  \bibfield  {author} {\bibinfo {author} {\bibfnamefont {H.}~\bibnamefont
  {Dutrieux}}, \bibinfo {author} {\bibfnamefont {C.}~\bibnamefont {Lorc\'e}},
  \bibinfo {author} {\bibfnamefont {H.}~\bibnamefont {Moutarde}}, \bibinfo
  {author} {\bibfnamefont {P.}~\bibnamefont {Sznajder}}, \bibinfo {author}
  {\bibfnamefont {A.}~\bibnamefont {Trawi\'nski}}, \ and\ \bibinfo {author}
  {\bibfnamefont {J.}~\bibnamefont {Wagner}},\ }\href {\doibase
  10.1140/epjc/s10052-021-09069-w} {\bibfield  {journal} {\bibinfo  {journal}
  {Eur. Phys. J. C}\ }\textbf {\bibinfo {volume} {81}},\ \bibinfo {pages} {300}
  (\bibinfo {year} {2021})},\ \Eprint {http://arxiv.org/abs/2101.03855}
  {arXiv:2101.03855 [hep-ph]} \BibitemShut {NoStop}%
\bibitem [{\citenamefont {Burkert}\ \emph {et~al.}(2021)\citenamefont
  {Burkert}, \citenamefont {Elouadrhiri},\ and\ \citenamefont
  {Girod}}]{Burkert:2021ith}%
  \BibitemOpen
  \bibfield  {author} {\bibinfo {author} {\bibfnamefont {V.~D.}\ \bibnamefont
  {Burkert}}, \bibinfo {author} {\bibfnamefont {L.}~\bibnamefont
  {Elouadrhiri}}, \ and\ \bibinfo {author} {\bibfnamefont {F.~X.}\ \bibnamefont
  {Girod}},\ }\href@noop {} {\  (\bibinfo {year} {2021})},\ \bibinfo {note}
  {[submitted to Nature Physics]},\ \Eprint {http://arxiv.org/abs/2104.02031}
  {arXiv:2104.02031 [nucl-ex]} \BibitemShut {NoStop}%
\bibitem [{\citenamefont {Shanahan}\ and\ \citenamefont
  {Detmold}(2019)}]{Shanahan:2018nnv}%
  \BibitemOpen
  \bibfield  {author} {\bibinfo {author} {\bibfnamefont {P.~E.}\ \bibnamefont
  {Shanahan}}\ and\ \bibinfo {author} {\bibfnamefont {W.}~\bibnamefont
  {Detmold}},\ }\href {\doibase 10.1103/PhysRevLett.122.072003} {\bibfield
  {journal} {\bibinfo  {journal} {Phys. Rev. Lett.}\ }\textbf {\bibinfo
  {volume} {122}},\ \bibinfo {pages} {072003} (\bibinfo {year} {2019})},\
  \Eprint {http://arxiv.org/abs/1810.07589} {arXiv:1810.07589 [nucl-th]}
  \BibitemShut {NoStop}%
\bibitem [{\citenamefont {Diehl}(2016)}]{Diehl:2015uka}%
  \BibitemOpen
  \bibfield  {author} {\bibinfo {author} {\bibfnamefont {M.}~\bibnamefont
  {Diehl}},\ }\href {\doibase 10.1140/epja/i2016-16149-3} {\bibfield  {journal}
  {\bibinfo  {journal} {Eur. Phys. J. A}\ }\textbf {\bibinfo {volume} {52}},\
  \bibinfo {pages} {149} (\bibinfo {year} {2016})},\ \Eprint
  {http://arxiv.org/abs/1512.01328} {arXiv:1512.01328 [hep-ph]} \BibitemShut
  {NoStop}%
\bibitem [{\citenamefont {Burkardt}(2003)}]{Burkardt:2002hr}%
  \BibitemOpen
  \bibfield  {author} {\bibinfo {author} {\bibfnamefont {M.}~\bibnamefont
  {Burkardt}},\ }\href {\doibase 10.1142/S0217751X03012370} {\bibfield
  {journal} {\bibinfo  {journal} {Int. J. Mod. Phys. A}\ }\textbf {\bibinfo
  {volume} {18}},\ \bibinfo {pages} {173} (\bibinfo {year} {2003})},\ \Eprint
  {http://arxiv.org/abs/hep-ph/0207047} {arXiv:hep-ph/0207047} \BibitemShut
  {NoStop}%
\bibitem [{\citenamefont {Miller}(2007)}]{Miller:2007uy}%
  \BibitemOpen
  \bibfield  {author} {\bibinfo {author} {\bibfnamefont {G.~A.}\ \bibnamefont
  {Miller}},\ }\href {\doibase 10.1103/PhysRevLett.99.112001} {\bibfield
  {journal} {\bibinfo  {journal} {Phys. Rev. Lett.}\ }\textbf {\bibinfo
  {volume} {99}},\ \bibinfo {pages} {112001} (\bibinfo {year} {2007})},\
  \Eprint {http://arxiv.org/abs/0705.2409} {arXiv:0705.2409 [nucl-th]}
  \BibitemShut {NoStop}%
\bibitem [{\citenamefont {Miller}(2009)}]{Miller:2009sg}%
  \BibitemOpen
  \bibfield  {author} {\bibinfo {author} {\bibfnamefont {G.~A.}\ \bibnamefont
  {Miller}},\ }\href {\doibase 10.1103/PhysRevC.80.045210} {\bibfield
  {journal} {\bibinfo  {journal} {Phys. Rev. C}\ }\textbf {\bibinfo {volume}
  {80}},\ \bibinfo {pages} {045210} (\bibinfo {year} {2009})},\ \Eprint
  {http://arxiv.org/abs/0908.1535} {arXiv:0908.1535 [nucl-th]} \BibitemShut
  {NoStop}%
\bibitem [{\citenamefont {Miller}(2019)}]{Miller:2018ybm}%
  \BibitemOpen
  \bibfield  {author} {\bibinfo {author} {\bibfnamefont {G.~A.}\ \bibnamefont
  {Miller}},\ }\href {\doibase 10.1103/PhysRevC.99.035202} {\bibfield
  {journal} {\bibinfo  {journal} {Phys. Rev. C}\ }\textbf {\bibinfo {volume}
  {99}},\ \bibinfo {pages} {035202} (\bibinfo {year} {2019})},\ \Eprint
  {http://arxiv.org/abs/1812.02714} {arXiv:1812.02714 [nucl-th]} \BibitemShut
  {NoStop}%
\bibitem [{\citenamefont {Jaffe}(2021)}]{Jaffe:2020ebz}%
  \BibitemOpen
  \bibfield  {author} {\bibinfo {author} {\bibfnamefont {R.~L.}\ \bibnamefont
  {Jaffe}},\ }\href {\doibase 10.1103/PhysRevD.103.016017} {\bibfield
  {journal} {\bibinfo  {journal} {Phys. Rev. D}\ }\textbf {\bibinfo {volume}
  {103}},\ \bibinfo {pages} {016017} (\bibinfo {year} {2021})},\ \Eprint
  {http://arxiv.org/abs/2010.15887} {arXiv:2010.15887 [hep-ph]} \BibitemShut
  {NoStop}%
\bibitem [{\citenamefont {Panteleeva}\ and\ \citenamefont
  {Polyakov}(2021)}]{Panteleeva:2021iip}%
  \BibitemOpen
  \bibfield  {author} {\bibinfo {author} {\bibfnamefont {J.~Y.}\ \bibnamefont
  {Panteleeva}}\ and\ \bibinfo {author} {\bibfnamefont {M.~V.}\ \bibnamefont
  {Polyakov}},\ }\href@noop {} {\  (\bibinfo {year} {2021})},\ \Eprint
  {http://arxiv.org/abs/2102.10902} {arXiv:2102.10902 [hep-ph]} \BibitemShut
  {NoStop}%
\bibitem [{\citenamefont {Kim}\ and\ \citenamefont {Kim}(2021)}]{Kim:2021jjf}%
  \BibitemOpen
  \bibfield  {author} {\bibinfo {author} {\bibfnamefont {J.-Y.}\ \bibnamefont
  {Kim}}\ and\ \bibinfo {author} {\bibfnamefont {H.-C.}\ \bibnamefont {Kim}},\
  }\href@noop {} {\  (\bibinfo {year} {2021})},\ \Eprint
  {http://arxiv.org/abs/2105.10279} {arXiv:2105.10279 [hep-ph]} \BibitemShut
  {NoStop}%
\bibitem [{\citenamefont {Brodsky}\ \emph {et~al.}(1998)\citenamefont
  {Brodsky}, \citenamefont {Pauli},\ and\ \citenamefont
  {Pinsky}}]{Brodsky:1997de}%
  \BibitemOpen
  \bibfield  {author} {\bibinfo {author} {\bibfnamefont {S.~J.}\ \bibnamefont
  {Brodsky}}, \bibinfo {author} {\bibfnamefont {H.-C.}\ \bibnamefont {Pauli}},
  \ and\ \bibinfo {author} {\bibfnamefont {S.~S.}\ \bibnamefont {Pinsky}},\
  }\href {\doibase 10.1016/S0370-1573(97)00089-6} {\bibfield  {journal}
  {\bibinfo  {journal} {Phys. Rept.}\ }\textbf {\bibinfo {volume} {301}},\
  \bibinfo {pages} {299} (\bibinfo {year} {1998})},\ \Eprint
  {http://arxiv.org/abs/hep-ph/9705477} {arXiv:hep-ph/9705477} \BibitemShut
  {NoStop}%
\bibitem [{\citenamefont {Weinberg}(1966)}]{Weinberg:1966jm}%
  \BibitemOpen
  \bibfield  {author} {\bibinfo {author} {\bibfnamefont {S.}~\bibnamefont
  {Weinberg}},\ }\href {\doibase 10.1103/PhysRev.150.1313} {\bibfield
  {journal} {\bibinfo  {journal} {Phys. Rev.}\ }\textbf {\bibinfo {volume}
  {150}},\ \bibinfo {pages} {1313} (\bibinfo {year} {1966})}\BibitemShut
  {NoStop}%
\bibitem [{\citenamefont {Gunion}\ \emph {et~al.}(1973)\citenamefont {Gunion},
  \citenamefont {Brodsky},\ and\ \citenamefont
  {Blankenbecler}}]{Gunion:1973ex}%
  \BibitemOpen
  \bibfield  {author} {\bibinfo {author} {\bibfnamefont {J.~F.}\ \bibnamefont
  {Gunion}}, \bibinfo {author} {\bibfnamefont {S.~J.}\ \bibnamefont {Brodsky}},
  \ and\ \bibinfo {author} {\bibfnamefont {R.}~\bibnamefont {Blankenbecler}},\
  }\href {\doibase 10.1103/PhysRevD.8.287} {\bibfield  {journal} {\bibinfo
  {journal} {Phys. Rev. D}\ }\textbf {\bibinfo {volume} {8}},\ \bibinfo {pages}
  {287} (\bibinfo {year} {1973})}\BibitemShut {NoStop}%
\bibitem [{\citenamefont {{Bracewell}}(2000)}]{Bracewell:2000abl}%
  \BibitemOpen
  \bibfield  {author} {\bibinfo {author} {\bibfnamefont {R.~N.}\ \bibnamefont
  {{Bracewell}}},\ }\href@noop {} {\emph {\bibinfo {title} {{The Fourier
  transform and its applications}}}}\ (\bibinfo  {publisher} {McGraw-Hill, New
  York},\ \bibinfo {year} {2000})\BibitemShut {NoStop}%
\bibitem [{\citenamefont {Rajan}\ \emph {et~al.}(2018)\citenamefont {Rajan},
  \citenamefont {Gorda}, \citenamefont {Liuti},\ and\ \citenamefont
  {Yagi}}]{Rajan:2018zzy}%
  \BibitemOpen
  \bibfield  {author} {\bibinfo {author} {\bibfnamefont {A.}~\bibnamefont
  {Rajan}}, \bibinfo {author} {\bibfnamefont {T.}~\bibnamefont {Gorda}},
  \bibinfo {author} {\bibfnamefont {S.}~\bibnamefont {Liuti}}, \ and\ \bibinfo
  {author} {\bibfnamefont {K.}~\bibnamefont {Yagi}},\ }\href@noop {} {\
  (\bibinfo {year} {2018})},\ \Eprint {http://arxiv.org/abs/1812.01479}
  {arXiv:1812.01479 [hep-ph]} \BibitemShut {NoStop}%
\bibitem [{\citenamefont {Miller}\ and\ \citenamefont
  {Brodsky}(2020)}]{Miller:2019ysh}%
  \BibitemOpen
  \bibfield  {author} {\bibinfo {author} {\bibfnamefont {G.~A.}\ \bibnamefont
  {Miller}}\ and\ \bibinfo {author} {\bibfnamefont {S.~J.}\ \bibnamefont
  {Brodsky}},\ }\href {\doibase 10.1103/PhysRevC.102.022201} {\bibfield
  {journal} {\bibinfo  {journal} {Phys. Rev. C}\ }\textbf {\bibinfo {volume}
  {102}},\ \bibinfo {pages} {022201} (\bibinfo {year} {2020})},\ \Eprint
  {http://arxiv.org/abs/1912.08911} {arXiv:1912.08911 [hep-ph]} \BibitemShut
  {NoStop}%
\bibitem [{\citenamefont {Freese}\ and\ \citenamefont
  {Miller}(2021{\natexlab{b}})}]{Freese:2021qtb}%
  \BibitemOpen
  \bibfield  {author} {\bibinfo {author} {\bibfnamefont {A.}~\bibnamefont
  {Freese}}\ and\ \bibinfo {author} {\bibfnamefont {G.~A.}\ \bibnamefont
  {Miller}},\ }\href@noop {} {\  (\bibinfo {year} {2021}{\natexlab{b}})},\
  \Eprint {http://arxiv.org/abs/2104.03213} {arXiv:2104.03213 [hep-ph]}
  \BibitemShut {NoStop}%
\bibitem [{\citenamefont {Carlson}\ and\ \citenamefont
  {Vanderhaeghen}(2009)}]{Carlson:2008zc}%
  \BibitemOpen
  \bibfield  {author} {\bibinfo {author} {\bibfnamefont {C.~E.}\ \bibnamefont
  {Carlson}}\ and\ \bibinfo {author} {\bibfnamefont {M.}~\bibnamefont
  {Vanderhaeghen}},\ }\href {\doibase 10.1140/epja/i2009-10800-0} {\bibfield
  {journal} {\bibinfo  {journal} {Eur. Phys. J. A}\ }\textbf {\bibinfo {volume}
  {41}},\ \bibinfo {pages} {1} (\bibinfo {year} {2009})},\ \Eprint
  {http://arxiv.org/abs/0807.4537} {arXiv:0807.4537 [hep-ph]} \BibitemShut
  {NoStop}%
\bibitem [{\citenamefont {Dirac}(1949)}]{Dirac:1949cp}%
  \BibitemOpen
  \bibfield  {author} {\bibinfo {author} {\bibfnamefont {P.~A.~M.}\
  \bibnamefont {Dirac}},\ }\href {\doibase 10.1103/RevModPhys.21.392}
  {\bibfield  {journal} {\bibinfo  {journal} {Rev. Mod. Phys.}\ }\textbf
  {\bibinfo {volume} {21}},\ \bibinfo {pages} {392} (\bibinfo {year}
  {1949})}\BibitemShut {NoStop}%
\bibitem [{\citenamefont {Newton}\ and\ \citenamefont
  {Wigner}(1949)}]{Newton:1949cq}%
  \BibitemOpen
  \bibfield  {author} {\bibinfo {author} {\bibfnamefont {T.~D.}\ \bibnamefont
  {Newton}}\ and\ \bibinfo {author} {\bibfnamefont {E.~P.}\ \bibnamefont
  {Wigner}},\ }\href {\doibase 10.1103/RevModPhys.21.400} {\bibfield  {journal}
  {\bibinfo  {journal} {Rev. Mod. Phys.}\ }\textbf {\bibinfo {volume} {21}},\
  \bibinfo {pages} {400} (\bibinfo {year} {1949})}\BibitemShut {NoStop}%
\bibitem [{\citenamefont {Kalnay}\ and\ \citenamefont
  {Toledo}(1967)}]{Kalnay:1967zz}%
  \BibitemOpen
  \bibfield  {author} {\bibinfo {author} {\bibfnamefont {A.~J.}\ \bibnamefont
  {Kalnay}}\ and\ \bibinfo {author} {\bibfnamefont {B.~P.}\ \bibnamefont
  {Toledo}},\ }\href {\doibase 10.1007/BF02721623} {\bibfield  {journal}
  {\bibinfo  {journal} {Nuovo Cim.}\ }\textbf {\bibinfo {volume} {48}},\
  \bibinfo {pages} {997} (\bibinfo {year} {1967})}\BibitemShut {NoStop}%
\bibitem [{\citenamefont {Pav\v{s}i\v{c}}(2018)}]{Pavsic:2018vbs}%
  \BibitemOpen
  \bibfield  {author} {\bibinfo {author} {\bibfnamefont {M.}~\bibnamefont
  {Pav\v{s}i\v{c}}},\ }\href {\doibase 10.1142/S0217732318501146} {\bibfield
  {journal} {\bibinfo  {journal} {Mod. Phys. Lett. A}\ }\textbf {\bibinfo
  {volume} {33}},\ \bibinfo {pages} {1850114} (\bibinfo {year} {2018})},\
  \Eprint {http://arxiv.org/abs/1804.03404} {arXiv:1804.03404 [hep-th]}
  \BibitemShut {NoStop}%
\bibitem [{\citenamefont {Sachs}(1962)}]{Sachs:1962zzc}%
  \BibitemOpen
  \bibfield  {author} {\bibinfo {author} {\bibfnamefont {R.}~\bibnamefont
  {Sachs}},\ }\href {\doibase 10.1103/PhysRev.126.2256} {\bibfield  {journal}
  {\bibinfo  {journal} {Phys. Rev.}\ }\textbf {\bibinfo {volume} {126}},\
  \bibinfo {pages} {2256} (\bibinfo {year} {1962})}\BibitemShut {NoStop}%
\bibitem [{\citenamefont {Hudson}\ and\ \citenamefont
  {Schweitzer}(2017)}]{Hudson:2017xug}%
  \BibitemOpen
  \bibfield  {author} {\bibinfo {author} {\bibfnamefont {J.}~\bibnamefont
  {Hudson}}\ and\ \bibinfo {author} {\bibfnamefont {P.}~\bibnamefont
  {Schweitzer}},\ }\href {\doibase 10.1103/PhysRevD.96.114013} {\bibfield
  {journal} {\bibinfo  {journal} {Phys. Rev. D}\ }\textbf {\bibinfo {volume}
  {96}},\ \bibinfo {pages} {114013} (\bibinfo {year} {2017})},\ \Eprint
  {http://arxiv.org/abs/1712.05316} {arXiv:1712.05316 [hep-ph]} \BibitemShut
  {NoStop}%
\bibitem [{\citenamefont {Brodsky}\ and\ \citenamefont
  {Lepage}(1989)}]{Brodsky:1989pv}%
  \BibitemOpen
  \bibfield  {author} {\bibinfo {author} {\bibfnamefont {S.~J.}\ \bibnamefont
  {Brodsky}}\ and\ \bibinfo {author} {\bibfnamefont {G.~P.}\ \bibnamefont
  {Lepage}},\ }\href {\doibase 10.1142/9789814503266_0002} {\bibfield
  {journal} {\bibinfo  {journal} {Adv. Ser. Direct. High Energy Phys.}\
  }\textbf {\bibinfo {volume} {5}},\ \bibinfo {pages} {93} (\bibinfo {year}
  {1989})}\BibitemShut {NoStop}%
\bibitem [{\citenamefont {Frankfurt}\ and\ \citenamefont
  {Strikman}(1981)}]{Frankfurt:1981mk}%
  \BibitemOpen
  \bibfield  {author} {\bibinfo {author} {\bibfnamefont {L.~L.}\ \bibnamefont
  {Frankfurt}}\ and\ \bibinfo {author} {\bibfnamefont {M.~I.}\ \bibnamefont
  {Strikman}},\ }\href {\doibase 10.1016/0370-1573(81)90129-0} {\bibfield
  {journal} {\bibinfo  {journal} {Phys. Rept.}\ }\textbf {\bibinfo {volume}
  {76}},\ \bibinfo {pages} {215} (\bibinfo {year} {1981})}\BibitemShut
  {NoStop}%
\bibitem [{\citenamefont {Bethe}\ and\ \citenamefont
  {Longmire}(1950)}]{PhysRev.77.647}%
  \BibitemOpen
  \bibfield  {author} {\bibinfo {author} {\bibfnamefont {H.~A.}\ \bibnamefont
  {Bethe}}\ and\ \bibinfo {author} {\bibfnamefont {C.}~\bibnamefont
  {Longmire}},\ }\href {\doibase 10.1103/PhysRev.77.647} {\bibfield  {journal}
  {\bibinfo  {journal} {Phys. Rev.}\ }\textbf {\bibinfo {volume} {77}},\
  \bibinfo {pages} {647} (\bibinfo {year} {1950})}\BibitemShut {NoStop}%
\bibitem [{\citenamefont {Kaplan}\ \emph {et~al.}(1999)\citenamefont {Kaplan},
  \citenamefont {Savage},\ and\ \citenamefont {Wise}}]{PhysRevC.59.617}%
  \BibitemOpen
  \bibfield  {author} {\bibinfo {author} {\bibfnamefont {D.~B.}\ \bibnamefont
  {Kaplan}}, \bibinfo {author} {\bibfnamefont {M.~J.}\ \bibnamefont {Savage}},
  \ and\ \bibinfo {author} {\bibfnamefont {M.~B.}\ \bibnamefont {Wise}},\
  }\href {\doibase 10.1103/PhysRevC.59.617} {\bibfield  {journal} {\bibinfo
  {journal} {Phys. Rev. C}\ }\textbf {\bibinfo {volume} {59}},\ \bibinfo
  {pages} {617} (\bibinfo {year} {1999})}\BibitemShut {NoStop}%
\bibitem [{\citenamefont {Chung}\ \emph {et~al.}(1988)\citenamefont {Chung},
  \citenamefont {Coester},\ and\ \citenamefont {Polyzou}}]{Chung:1988mu}%
  \BibitemOpen
  \bibfield  {author} {\bibinfo {author} {\bibfnamefont {P.~L.}\ \bibnamefont
  {Chung}}, \bibinfo {author} {\bibfnamefont {F.}~\bibnamefont {Coester}}, \
  and\ \bibinfo {author} {\bibfnamefont {W.~N.}\ \bibnamefont {Polyzou}},\
  }\href {\doibase 10.1016/0370-2693(88)90995-1} {\bibfield  {journal}
  {\bibinfo  {journal} {Phys. Lett. B}\ }\textbf {\bibinfo {volume} {205}},\
  \bibinfo {pages} {545} (\bibinfo {year} {1988})}\BibitemShut {NoStop}%
\bibitem [{\citenamefont {Hudson}\ and\ \citenamefont
  {Schweitzer}(2018)}]{Hudson:2017oul}%
  \BibitemOpen
  \bibfield  {author} {\bibinfo {author} {\bibfnamefont {J.}~\bibnamefont
  {Hudson}}\ and\ \bibinfo {author} {\bibfnamefont {P.}~\bibnamefont
  {Schweitzer}},\ }\href {\doibase 10.1103/PhysRevD.97.056003} {\bibfield
  {journal} {\bibinfo  {journal} {Phys. Rev. D}\ }\textbf {\bibinfo {volume}
  {97}},\ \bibinfo {pages} {056003} (\bibinfo {year} {2018})},\ \Eprint
  {http://arxiv.org/abs/1712.05317} {arXiv:1712.05317 [hep-ph]} \BibitemShut
  {NoStop}%
\bibitem [{\citenamefont {Freese}\ and\ \citenamefont
  {Clo\"et}(2019)}]{Freese:2019bhb}%
  \BibitemOpen
  \bibfield  {author} {\bibinfo {author} {\bibfnamefont {A.}~\bibnamefont
  {Freese}}\ and\ \bibinfo {author} {\bibfnamefont {I.~C.}\ \bibnamefont
  {Clo\"et}},\ }\href {\doibase 10.1103/PhysRevC.100.015201} {\bibfield
  {journal} {\bibinfo  {journal} {Phys. Rev. C}\ }\textbf {\bibinfo {volume}
  {100}},\ \bibinfo {pages} {015201} (\bibinfo {year} {2019})},\ \Eprint
  {http://arxiv.org/abs/1903.09222} {arXiv:1903.09222 [nucl-th]} \BibitemShut
  {NoStop}%
\end{thebibliography}%

\end{document}